\documentclass[trackchanges]{aastex701}

\usepackage{gensymb}
\usepackage{rotating}
\usepackage{lineno}
\usepackage{hyperref}

\begin{document}

\title{3I/ATLAS (C/2025 N1): Direct Spacecraft Exploration of a Possible Relic of 
\\Planetary Formation at ``Cosmic Noon''}

\correspondingauthor{T. Marshall Eubanks}

\author[0000-0001-9543-0414]{T. Marshall Eubanks}
\affiliation{Space Initiatives Inc, Princeton, WV 24740, USA}
\email[show]{tme@space-initiatives.com}

\author[0000-0002-1978-8243]{Bruce G. Bills}
\affiliation{Jet Propulsion Laboratory, 4800 Oak Grove Drive, Pasadena, CA 91109, USA}
\email{Bruce.Bills@jpl.nasa.gov}

\author[0000-0003-1116-576X]{Adam Hibberd}
\affiliation{Institute for Interstellar Studies (i4is US), Oak Ridge, TN 37830, USA}
\email{hibberdadam994@gmail.com}

\author{W. Paul Blase}
\affiliation{Space Initiatives Inc, Princeton, WV 24740, USA}
\email{wpb@space-initiatives.com}

\author[0000-0003-1763-6892]{Andreas M. Hein}
\affiliation{SnT, University of Luxembourg,  L-4365 Luxembourg}
\email{andreas.hein@i4is.org}

\author[0000-0003-3924-7935]{Robert G. Kennedy III}
\affiliation{Institute for Interstellar Studies (i4is US), Oak Ridge, TN 37830, USA}
\email{robert.kennedy@i4is.org}

\author[0000-0003-1322-7485]{Adrien Coffinet}
\affiliation{Coulommiers, France}
\email{adrien.coffinet2@gmail.com}

\author[0000-0002-3136-6537]{Jean Schneider}
\affiliation{Observatoire de Paris - LUTH, 92190 Meudon, France}
\email{Jean.Schneider@obspm.fr}

\author[0000-0003-0626-1749]{Pierre Kervella}
\affiliation{LIRA, Observatoire de Paris, Universit\'e PSL, Sorbonne Universit\'e, Universit\'e Paris Cit\'e, CY Cergy Paris Universit\'e, CNRS, 5 place Jules Janssen, 92195 Meudon, France}
\affiliation{French-Chilean Laboratory for Astronomy, IRL 3386, CNRS and U. de Chile, Casilla 36-D, Santiago, Chile}
\email{pierre.kervella@obspm.fr}
\author[0009-0007-3343-2001]{Carlos Gomez de Olea Ballester}
\affiliation{TU Munich, D-80333 Munich, Germany}
\email{carlos.olea@tum.de}

\begin{abstract}
The interstellar object 3I/ATLAS (also  C/2025 N1 (ATLAS), henceforth, 3I), discovered by the  ATLAS Chile telescope on 2025 July 1, was rapidly  revealed to be the third known interstellar object (ISO) transiting the solar system, with an incoming velocity at infinity 
of 57.9763 $\pm$  0.0044 km s$^{-1}$. An examination of 3I's pre-encounter kinematics shows that it is likely to be  an object from the galactic thick disk, and thus a remnant of the Galaxy's ``cosmic noon'' period of intense star formation $\sim$9 - 13 gigayears ago. This kinematic assignment of 3I to the thick disk can be tested observationally in the  transit of 3I through the solar system; if confirmed, 3I will provide a means to explore the stellar and planetesimal formation process, and its astrobiological implications, in an early period of galactic history.  
Unfortunately for terrestrial observers, the 3I perihelion will happen when it
is on the other side of the Sun as seen from Earth, at a solar elongation of 12.80$\degree$, rendering observation  from Earth (or near-Earth space telescopes)  hard or impossible. With a retrograde orbit inclined 175.114$\degree$ (only 4.886$\degree$ from the ecliptic plane), and a trajectory passing inside the orbit of Mars, 3I will pass relatively close to a number of already launched interplanetary spacecraft. We examined the observational opportunities of 15 spacecraft, both those on heliocentric orbits and also the spacecraft grouping at the planet Mars, and find a strong science case for observations in the periods of the close approaches of the Psyche spacecraft on 2025 September 4, at 0.302 AU, the martian spacecraft array on 2025 October 3, and the Juice spacecraft on 2025 November 4. 
In addition, the Europa Clipper, Hera and even the more distant Lucy spacecraft may pass through 3I's cometary tail in the period after its perihelion passage, potentially directly observing the conditions and composition there. 
Finally, three heliophysics space observatories (SOHO, Solar Orbiter and the Parker Solar Probe) will have 3I pass through their instrument's fields of view (FOV) in this period;  the Parker Solar Probe and even the solar coronagraphs  may be able to monitor 3I at intervals in the period from late September through mid November in 2025.
Spacecraft observations could, to the extent they are possible,   provide the only source of spectral and imaging data during  the 3I perihelion passage,  constrain the phase angle dependence of 3I coma dust, and test the 3I thick disk origin hypothesis.  \end{abstract}



\tableofcontents

\section{Introduction} 
\label{sec:introduction}

The recent discovery of Interstellar Objects (ISOs)
passing through the solar system 
on clearly hyperbolic orbits
opens the potential for direct observation of material from other star systems, both telescopically, from on and near the Earth
(see, e.g., \citep{Trilling-et-al-2018-b,Guzik-et-al-2019-a}, and potentially also with flyby, rendezvous and sample return spacecraft missions \citep{Hein-et-al-2017-a,Seligman-Laughlin-2018-a,Moore-et-al-2020-a,Hein-et-al-2020-b,Eubanks-et-al-2020-b}.

The first confirmed ISO, 1I/'Oumuamua (henceforth 1I) was discovered on
2017 October 19, and remains fairly mysterious, as it was detected after its perihelion passage and observations were only possible through 2018 January 2, as it moved away from the Sun. In many ways, 1I seems to have been a small interstellar asteroid, with an absolute visual magnitude, H, = 22, and no detection of either a cometary coma or any gaseous spectral emission lines, although orbital analysis did reveal a significant non-gravitational acceleration \citep{Micheli-et-al-2018-a}, indicating either some otherwise undetected outgassing \citep{Seligman-et-al-2019-a} or a very low mass/area ratio \citep{Bialy-Loeb-2018-a}. 

The second confirmed ISO, 2I/Borisov (henceforth, 2I), was discovered by the amateur astronomer Gennadiy  Borisov on 2019 August 30 \citep{Guzik-et-al-2019-a} at a distance of 3 astronomical units (AU) from the Sun and a solar elongation (Sun-observer-target angle) of only  37.238$\degree$. During its solar system passage 2I was an active object  superficially similar to a small solar system comet \citep{Jewitt-and-Luu-2019-a},  with  an estimated nucleus
diameter of order 1 km \citep{Jewitt-et-al-2019-a}. It exhibited typical comet spectral features, enabling compositional studies which showed it to be  carbon-depleted and relatively nitrogen-rich compared to solar system comets \citep{Deam-et-al-2025-a}. 2I  also experienced a fragmentation event, losing a relatively small ``boulder-sized'' component in 2020 March,  when it was $\sim$3 AU from the Sun \citep{Bolin-et-al-2020-a,Jewitt-et-al-2020-a}.

3I was discovered at a distance of 4.52 AU from the Sun, just inside the orbit of Jupiter \citep{Bolin-et-al-2025-a}. 
and was rapidly revealed to be both an interstellar object and an active comet, with a 
dusty coma  $\sim$25,000 km in diameter
\citep{de_la_Fuente_Marcos-et-al-2025-a}, with 
cometary activity being discovered in precovery data acquired when it was at a distance of 4.85 AU from the Sun, on 2025 June 21, \citep{Chandler-et-al-2025-a}. 
Although 3I has a red spectrum \citep{Opitom-et-al-2025-a}, 
and there is a claimed detection of water ice spectral reflection features in the  Infrared (IR) \citep{Yang-et-al-2025-a}, there are no reports as yet of gas emission lines from 3I \citep{Alvarez-Candal-et-al-2025-a}. 
3I passed within 3 AU of the Sun on  2025 August 16; after that time its activity and spectra can be compared directly with that of 2I on its inbound trajectory. 

Estimates of the size of the 3I nucleus are at present still upper bounds, which have decreased with time. The most recent estimate is from HST observations, which yield  a radius, R, 
0.16 $\le$ R $\le$ 2.8 km, corresponding to an absolute magnitude, H, 
range of 23.1 $\ge$ H $\ge$ 15.4 \citep{Jewitt-et-al-2025-a}, all assuming a ``red'' geometric albedo of 0.04. Ground-based observations of 3I provide a rotation period of  is 16.79 $\pm$ 0.23 h \citep{de_la_Fuente_Marcos-et-al-2025-a}, which renders any rotational disruption unlikely. 

The purpose of this work is to encourage scientific  observations of 3I by showing what observations could be possible, and what their scientific benefit could be. Of course, spacecraft mission operation and science teams will have to decide whether and when their spacecraft can safely and usefully support observations of 3I.

\section{Dynamical Origins of Interstellar Objects} \label{sec:dynamical-origins}

The vector velocity of any  object orbiting in the Galaxy provides information about its past history. This information can be obtained for an ISO from an analysis of astrometric observations of its passage through the solar system, which provide its scalar \textit{a priori} velocity at infinity, v$_{\infty}$, from the hyperbolic semi-major axis of the incoming object, and the velocity unit vector from the projection of the object's position and velocity into the distant past. (Henceforth the ISO velocities discussed will exclusively refer to velocities from before their interaction with the solar system.) 

Stellar perturbations make it hard to predict the detailed galactic trajectories of
ISOs over intervals much longer than a few million years \citep{Zhang-2017-a},
less  than 10$^{-3}$ of the history of the galaxy, and no firm association with a nearby star has been claimed for either 1I or 2I \citep{Bailer-Jones-et-al-2019-a,Hallatt-Wiegert-et-al-2019-a}, nor is one likely for 3I.
However, velocities of thin disk stars orbiting relatively near the Sun  are highly non-uniform \citep{Ramos-et-al-2018-a},
with a substantial fraction of the stars in the solar neighborhood being concentrated
in dynamical streams (also known as associations or moving groups) moving in the galaxy
(see, e.g., \citep{Kushniruk-et-al-2017-a,Gaia-et-al-2018-b}).
Even without finding the natal star system of an ISO, therefore, 
association with a stellar stream can serve as a proxy to provide bounds on the age, metallicity, and chemical composition of the source stellar system.

\subsection{Kinematics of ISOs and Nearby Stellar Systems}
\label{subsec:ISO-Kinematics}

In our current 3I orbital solution using \textbf{find\_orb},
2611 of 2793 observations with data from  177 observatories were used from the period
2025 May 22-August 12, with root mean square right ascension and declination residuals of 0.521 and 0.626 seconds of arc (''), respectively. The 
v$_{\infty}$ from this solution was 57.9763 $\pm$  0.0044 km s$^{-1}$; solutions with different datasets, editing  and parameterizations all have v$_{\infty}$ within a few 10's of m s$^{-1}$ of 57.98 km s$^{-1}$. 

The resulting velocity vectors for 3I in the distant past (in practice 1600 Jan 1 is sufficient), together with similar velocity estimates for 1I and 2I \citep{Eubanks-2019-b}, were converted into galactic UVW coordinate components for an analysis of the galactic kinematics of these ISOs. (In this paper we use galactic U, V and W velocity components in a right-hand system with unit vectors pointing towards the galactic center for U, in the direction of galactic rotation for V, and towards the galactic North pole for W, respectively.)

\begin{figure*}[ht!]
\plotone{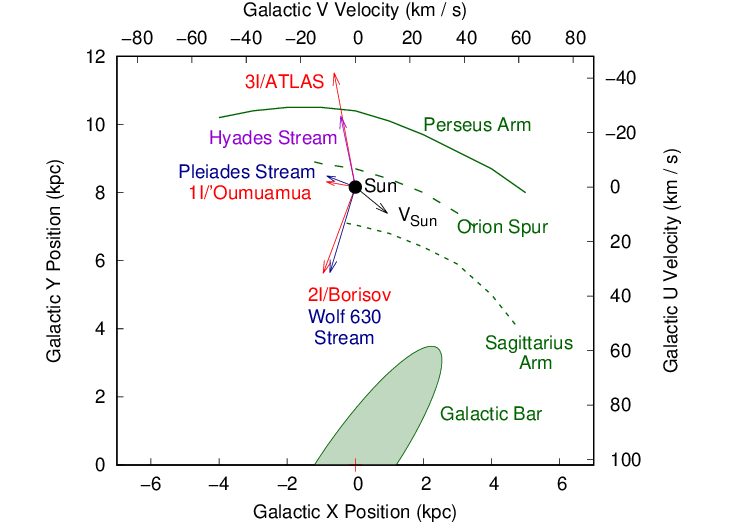}
\caption{A galactic view of ISO and stellar kinematics.
The equatorial (U and V) velocities of the 3 known ISOs 
compared with a simplified map of galactic structure. 1I and 2I velocity's match well with known streams, but 
3I does not match the galactic velocities of any major stellar stream. Note that 2I was, at the time of its encounter with the solar system, moving from the outer parts of the galaxy towards its perigalacticon, while 3I was moving away from its perigalacticon  into the outer galaxy.
\label{fig:Milky-way-map-1}}
\end{figure*}

Figure \ref{fig:Milky-way-map-1} shows the U and V galactic barycentric velocity coordinates for all three known ISOs together with the in-plane velocity vectors for some well-known stellar streams, and a rough sketch of some major features on this side of the galaxy.
This  shows  that the incoming 1I and 2I  velocity vectors are close to the motion of the Pleiades and Wolf 630 dynamical streams, respectively,  in U and V (galactic plane) velocity components, suggesting that (before their interaction with the solar system)  1I was a member of the Pleiades dynamical stream
\citep{Feng-Jones-2018-a,Eubanks-2019-a,Eubanks-2019-b}
and  2I  was a member of the Wolf 630 stream \citep{Eubanks-2019-c}.
However, while  3I and the Hyades Stream \citep{Famaey-et-al-2008-a} galactic plane motions are aligned in Figure \ref{fig:Milky-way-map-1}, their velocity amplitudes differ, and 3I  and the Hyades stream also have considerably different motions in W, out of the galactic plane.

The 1I and 2I galactic velocities are in general near the centroids of the  velocities of the Pleiades and Wolf 630 streams, respectively, suggesting that these objects were members of these streams before their encounters with the solar system. The Pleiades stream   consists of stars with ages between roughly 10$^{7}$ and 10$^{9}$ years  \citep{Chereul-et-al-1998-b,Famaey-et-al-2008-a,Bovy-Hogg-2009-a}, suggesting that
1I is a relatively young object. 
Antoja \textit{et al}
\citep{Antoja-et-al-2008-a} determined the kinematics of
over 16,000 nearby stars with known ages; the Wolf 630
stream (their moving group 17) has in their data almost no stars younger than
 0.5 Gyr, and are at most 2 - 8 Gyr old, suggesting that 2I/Borisov may be roughly the age of the solar system, albeit with  a large dating uncertainty. 
 By contrast, 3I does not appear to be part of any stellar stream; determining its kinematic origin will require a consideration of  the broader galactic structure.

Table \ref{table:Stream-kinematics}  shows the results of a statistical comparison of all three U, V, W velocity components of the known ISOs and their closest matching stellar streams. 
These  stream velocity estimates are averages of estimates from  \cite{Kushniruk-et-al-2017-a,Chereul-et-al-1998-b,Liang-et-al-2017-a,Gaia-et-al-2018-b,Famaey-et-al-2008-a}, and the stream velocity formal errors are based on the scatter of the available estimates.
While 1I and the Pleiades stream kinematics  agree statistically very well, and the agreement between the 2I and the Wolf 630 stream velocities is also  acceptable, the 3I and Hyades stream velocities are quite different. 
We thus assert that the suggested association with 3I is not real. 

\begin{table}[ht]
        \centering
        \begin{tabular}{ c c c c c} 
          \hline
            & 
           \multicolumn{3}{ c }{Solar Frame Velocity}   \\
           & U & V & W &  $\chi^{2}$(3 d.f.)\\
            & km s$^{-1}$ & km s$^{-1}$ & km s$^{-1}$\\
        \hline
1I  &   -11.58 $\pm$ 0.1  &  -22.43 $\pm$ 0.1 & -7.86 $\pm$ 0.1  \\
Pleiades  &   -13.7 $\pm$ 3.8  &  -22.3 $\pm$ 1.4 & -8.3 $\pm$ 2.0 \\
Difference  &   2.1 $\pm$ 3.8  &  -0.1 $\pm$ 1.4 & -2.5 $\pm$ 2.0 & 0.37\\
          \hline
2I  &   22.03 $\pm$ 0.07  &  -23.64 $\pm$ 0.07 & 1.03 $\pm$ 0.07  \\
Wolf 630  &   21.8 $\pm$ 3.9  &  -21.1  $\pm$ 4.9 & -12.1 $\pm$ 7.8 \\
Difference  &   0.3 $\pm$ 3.9  &  -2.5  $\pm$ 4.9 & 13.1 $\pm$ 7.8 & 3.12\\
          \hline
3I  &   -51.57 $\pm$ 0.50  &  -19.59 $\pm$ 0.50 & 19.18 $\pm$ 0.50  \\
Hyades  &   -35.7 $\pm$ 4.0  &  -17.2 $\pm$ 2.5 & -10.3 $\pm$ 8.2 \\
Difference  &   -15.9 $\pm$ 4.0  &  -2.4 $\pm$ 2.6 & 29.5 $\pm$ 8.2 & 29.27\\
          \hline
       \end{tabular}
    \caption{\textbf{Galactic Stream Kinematics} 
    U, V, W velocity coordinates for the 3 ISOs and their best-matching galactic streams.
    The 1I and 2I velocities are reasonably close matches to  Pleiades and Wolf 630 stream velocities, respectively, while the 3I velocity is not consistent with the Hyades stream velocity in both the U and W components. Note that these velocities are in the solar system frame, and are not corrected to the Local Standard of Rest (LSR). 
    }
\label{table:Stream-kinematics}
\end{table}
 
\subsection{Thin or Thick Disk}
\label{subsec:thin-or-thick}

The Sun and its planetary system 
reside in the galactic thin disk, with orbits within a few hundred pc of the galactic plane and out of plane (galactic W) velocities $\lesssim$15 km s$^{-1}$ \citep{Vieira-et-al-2022-a}. By using velocity measurements of stars near the Sun, we can better characterize the populations of stars in various galactic associations, including both the stellar streams and the galactic disks. 

We used the Gaia EDR-3 Catalogue of Nearby Stars (GCNS), a nearly complete catalog of stars within 100 parsecs of the solar system \citep{Gaia-et-al-2020-b},  to determine the kinematic structure of the local Galaxy. 
The 331,312 stars in the GCNS  are thought to include at least 92\% of stars of stellar type M9 or earlier within 100\,pc of the Sun, providing a nearly complete catalog of stars within the solar neighborhood \citep{Gaia-et-al-2020-b}. However, due to a lack of radial velocities for some of those stars we could only use a total of 77,132 stars from this catalog, those stars having the radial velocity  data needed to provide all three components of 3-D velocity, and also passing two catalog quality checks, requiring the ``probability of having reliable astrometry'' to be $\ge$ 0.75 and the ``maximum  renormalized unit weight error'' (RUWE) to be $\le$ 10.  
 
\begin{table}[ht]
        \centering
        \begin{tabular}{ c c c c c} 
          \hline
            & 
           \multicolumn{3}{ c }{Galactic Frame Velocity} \\
           & U & V & W &  $\chi^{2}$(3 d.f.)\\
            & km s$^{-1}$ & km s$^{-1}$ & km s$^{-1}$\\
        \hline
Thin Disk &   1 $\pm$ 31   &  239 $\pm$  20 & 0 $\pm$ 11  \\
Thick Disk & -1 $\pm$ 49  &  225 $\pm$   35 & 0 $\pm$ 22  \\
          \hline
1I  &   -1.93 $\pm$ 2.7  &  222.96 $\pm$ 2.0 & -0.91 $\pm$ 1.20  \\
- Thin Disk  &   -2.9 $\pm$ 31  &  -16.0 $\pm$ 20 & -0.9 $\pm$ 11 & 0.65\\
- Thick Disk  &   -0.9 $\pm$ 49  &  -2.0 $\pm$ 35 & -0.9 $\pm$ 33 & 0.01\\
          \hline
2I  &   31.68 $\pm$ 2.7  &  221.75 $\pm$ 2.0 & 7.98 $\pm$ 1.20  \\
- Thin Disk  &   30.7 $\pm$ 31  &  -17.3  $\pm$ 20 & 8.0 $\pm$ 11 & 2.23\\
- Thick Disk  &  32.7 $\pm$ 49  &  -3.5 $\pm$ 35 & 8.0 $\pm$ 33 & 0.51\\ 
          \hline
3I  &   -42.92 $\pm$ 2.7  &  225.80 $\pm$ 2.0 & 26.15 $\pm$ 1.20  \\
- Thin Disk  &   -40.9 $\pm$ 31  &  -13.2 $\pm$ 20 & 26.1 $\pm$ 11 & 7.90\\
- Thick Disk  &   -40.9 $\pm$ 49  &  0.8 $\pm$ 35 &  26.1 $\pm$ 33 & 1.32\\ 
          \hline
       \end{tabular}
    \caption{\textbf{Milky Way Disk Kinematics} 
   U, V, W velocity coordinates in the galactic frame. The ISO velocities have been adjusted to this frame using  the LSR velocity of \cite{Eubanks-et-al-2021-a}  and the circular rotation velocity of \cite{Mroz-et-al-2019-a}). 
    The thin and thick disk mean velocities and scatters are from \cite{Vieira-et-al-2022-a}, after conversion of those velocities to our right-handed UVW system. 
    This statistical comparison shows that, while the
    3I galactic velocity is statistically consistent with the thick disk velocity, it is not consistent with the thin disk velocity estimate and scatter at the 95\% significance level.
}
\label{table:Disk-kinematics}
\end{table}

Table \ref{table:Disk-kinematics} shows a statistical analysis, similar to that in 
Table \ref{table:Stream-kinematics}, 
of a comparison of UVW ISO velocities and the thick and thin disk velocity models provided by \citep{Vieira-et-al-2022-a}. In this comparison the LSR velocity estimated in \citep{Eubanks-et-al-2021-a}
and the circular rotation speed (in V) of 233.6 km s$^{-1}$ of \cite{Mroz-et-al-2019-a} were added to the ISO velocities. Note that the rms scatters provided for the disk velocity models do not reflect measurement errors, but the scatter in the velocities of the relevant disk populations. 

The mean values of the U and W (polar) velocities are near zero for both the thin and the thick disk; the magnitudes of these velocity components thus indicates which disk an ISO comes from. The biggest mean component difference is that the thick disk in the solar neighborhood rotates a little slower about the galaxy, with a V component 14 km s$^{-1}$ smaller than for the thin disk model. Interestingly all 3 known ISOs appear to share this slower galactic rotation, but the scatter in V velocities is larger than this difference; 
the significance of this bias will not be known until more ISOs have been discovered and their velocities analyzed. 

The statistics presented in Table \ref{table:Disk-kinematics} show that the ISO 3I does not fit the thin disk model with a $
\chi^{2}$ for 3 degrees of freedom is 7.90, a probability limit of 95\%. This is a weak, but positive, confirmation of 3I being a member of the thick disk. 
Other researchers come to a variety of conclusions 
from the 3I kinematic data. 
\cite{Hopkins-et-al-2025-a} conclude that 3I is in the thick disk, while 
\cite{Taylor-Seligman-2025-a} conclude that 3I is approximately 3-11 Gyr old, and thus that is ''likely originated in the thick disk of the Galaxy, providing a sample of system formation in this region.''
Conversely, \cite{de_la_Fuente_Marcos-et-al-2025-a} use the kinematic data to select a sample of kinematic analog stars from Gaia DR3 and obtain an estimate of [Fe/H] = -0.04 $\pm$ 0.14, which places 3I in the thin disk. 

Although 3I has been assigned to the thick disk on kinematic grounds, a halo origin remains a possible explanation and could account for the observed discrepancies in the out-of-plane kinematic components discussed in Section \ref{subsec:ISO-Kinematics}. Such a scenario is unlikely but not impossible, as 3I could either be a halo object by birth or have been scattered into a colder galactic orbit following a past stellar encounter. Given the low metallicity and enhanced $alpha$-element abundances characteristic of halo stars \citep{Helmi-2008-a}, a halo origin would make 3I a particularly intriguing case, potentially even a relic of  Population III star formation. 
The upcoming 3I measurement campaign  offers an opportunity to directly confront the thick disk 3I hypothesis with observations, by attempting to observe characteristic signatures of an origin in that disk. 

\subsection{The Galactic Context}
\label{subsec:galactic-context}

\begin{figure*}[ht!]
\epsscale{1.0}
\plotone{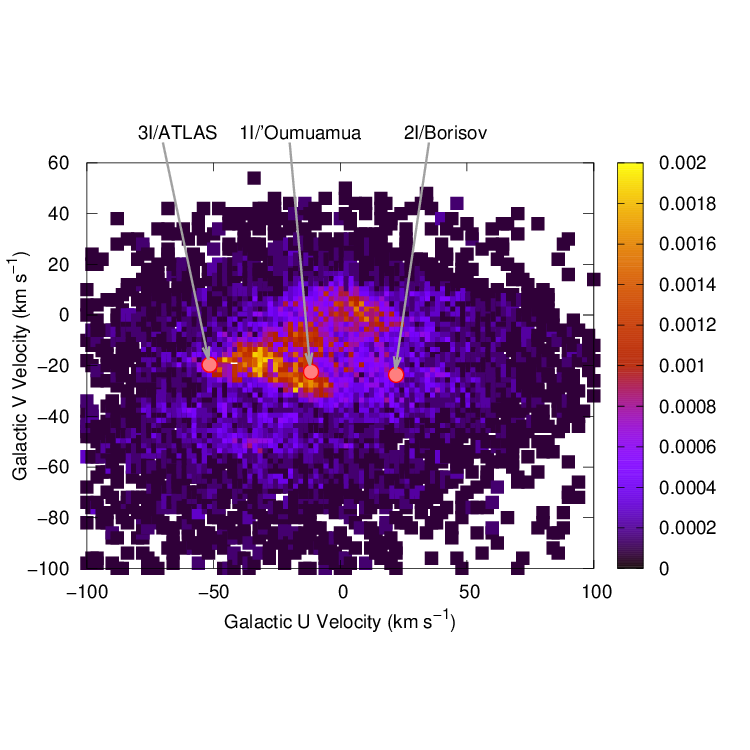}
\caption{The relative density of GCNS stars in terms of galactic U and V velocities. These velocities are all in the solar frame, with the LSR being at (-9.65, -11.79) km s$^{-1}$, slightly above the icon for 1I. 
\label{fig:Table1_GCNS_UV_heat.9_close}}
\end{figure*}

\begin{figure*}[ht!]
\epsscale{1.0}
\plotone{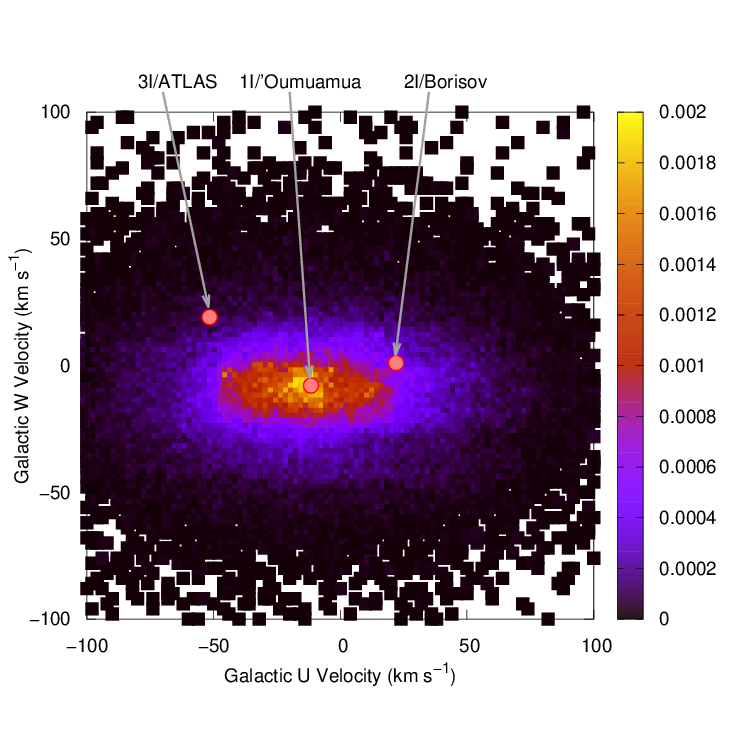}
\caption{The relative density of GCNS stars in terms of galactic U and W velocities. 
These velocities are all in the solar frame, with the LSR being at (-9.65, -6.95) km s$^{-1}$, close to the icon for 1I. 
\label{fig:Table1_GCNS_UW_heat.7}}
\end{figure*}

Figures \ref{fig:Table1_GCNS_UV_heat.9_close} and \ref{fig:Table1_GCNS_UW_heat.7} show 2-dimensional histograms of the density of stars in the EDR-3 catalog (edited as described above) for the UV and UW velocity components, respectively, together with labeled disks showing the velocities of the 3 ISOs. (These plots are all in the solar system frame and none of the velocities have been adjusted to the LSR.) The 
density peaks  in Figure  \ref{fig:Table1_GCNS_UV_heat.9_close}
seem to mostly represent known stellar / dynamical streams (such as the Pleiades stream) \citep{Hunt-Vasiliev-2025-a,Ramos-et-al-2018-a,Vieira-et-al-2022-a}, with these density peaks being embedded 
in larger arch-like kinematic structures \citep{Gaia-et-al-2018-b}, which seem to be related to the dynamical galactic bar resonances \citep{Khoperskov-et-al-2020-a}. 

While it might seem in Figure \ref{fig:Table1_GCNS_UV_heat.9_close}  that 3I is in a density peak related to the thin disk, Figure  \ref{fig:Table1_GCNS_UW_heat.7} shows that 3I's velocity is not near the thin disk's 3-D distribution at all. The ISO 1I appears to be  kinematically in the thin disk; 2I may be a mixed case, but it kinematically seems to be quite consistent with an origin in the thin disk. 

\begin{table}[ht]
        \centering
        \begin{tabular}{ c  c  c c c c c} 
        \hline
        ISO  & a  & 
           e     & i &  q  & Q & a sin(i)\\
        1I & 7429 pc &  0.0894 & 0.2338$\degree$ & 6765 pc &  8093 pc & 30.3 pc \\
        2I & 7496 pc &  0.1617 & 2.0610$\degree$ & 6284 pc &  8707 pc & 270 pc \\
        3I & 7924 pc &  0.1825 & 6.6024$\degree$ & 6477 pc &  9370 pc & 911 pc \\

        \hline
        \end{tabular}
\caption{\textbf{Keplerian Galactic Orbits.} 
    Osculating orbit elements for the 3 ISOs based on their position at the Sun (assumed to be at a distance of 8090 pc from the galactic disk \citep{Mroz-et-al-2019-a})
    and their velocity vectors as given in Table \ref{table:Disk-kinematics}. ``q'' and ``Q'' are the  perigalacticon and apogalacticon distances, respectively. Note that actual galactic orbits can be highly non-keplerian and so these osculating orbital estimates are not necessarily good models for their full galactic orbits.
    \label{table:galactic-orbits}
    }
\end{table}

Table \ref{table:galactic-orbits} shows the Keplerian osculating orbital elements for the ISOs at the time of their encounter with the solar system (of course, these elements will not apply to their post solar encounter orbits). 1I appears to have been discovered very near its apogalacticon but otherwise had a nearly circular  low-inclination galactic orbit. The thin and thick disks can be fit with exponential scale heights out of the galactic plane of $\sim$300 and 900 pc, respectively
\citep{Juric-et-al-2008-a}. While the a sin(i) of 1I is well within the thin disk scale height, the orbit of 2I takes it almost out to the thin disk scale height and 3I's orbit takes it almost a kpc away from the galactic plane, 3 scale heights for the thin disk, but almost at the thick disk scale height, fully consistent with a thick disk origin for this object. 

Interestingly, the orbits of both 2I and 3I appear to be connected to the  main resonances of the Milky Way bar shown in Figure \ref{fig:Milky-way-map-1}, with 2I's perigalacticon being near  the bar corotation resonance at
$\sim$6.2 kpc and 3I's apogalacticon being at or beyond the outer Lindblad resonance at $\sim$9 kpc; the orbit of these objects thus appear to share the imprint of the Milky
Way bar and spiral density waves on the  disk population, suggesting that there might be ISO streams based, not on their formation region, but on galactic resonances.

\subsection{The Composition of Thick Disk Planetesimals}
\label{subsec:thick disk-composition}

Planets, and thus presumably the planetesimals and ISOs that formed with them, will be imprinted by the conditions of their formation, and also by the material available to form them.
The thick disk is old, with star-formation peaks between 12 and 10 Gyr ago and steadily increasing metallicities (ending with some stars at super-solar values
\citep{Fernandez-Alvar-et-al-2025-a}.
$\alpha$-elements are produced mainly during Type II Supernova (SNII) explosions of massive stars, which occur relatively rapidly (in $\sim$ 10$^{7}$ years), whereas iron is also produced on a much longer time-scale ($\sim$10$^{9}$ years) by Type Ia SNe of less massive stars 
\citep{Kordopatis-et-al-2011-a}.
Heuristically, star formation in the thick disk simply did not last long enough for enough iron and other iron-peak elements (Cr, Mn, Fe, Co and Ni) to be made to solar (i.e., thin disk) abundances.

Compared to stars of the thin disk, those from the thick
disk thus tend to be older, iron and iron-peak poor and with $\alpha$-elements (C, O, Ne, Mg, Si, and S) enhanced compared to iron \citep{Bashi-Zucker-2022-a},
with -0.3 $<$ [Fe/H] $<$ -1.2
\citep{Reddy-et-al-2006-a,Snaith-et-al-2014-a}
 and 0.1 $<$ [$\alpha$/Fe] $<$ 0.45 (although [$\alpha$/Fe] decreases as the metallicity increases) \citep{Recio-Blanco-et-al-2014-a}.
It is thus not surprising that rocky exoplanets from the thick disk are found to have lower densities, presumably indicative of lower iron abundances and smaller iron cores
\citep{Weeks-et-al-2025-a}. If smaller iron cores cause them to  have smaller or missing core-dynamo magnetic fields, these planets may also be less habitable \citep{Lazio-et-al-2017-a}.

Thick disk ISOs should thus be enhanced in $\alpha$-elements and depleted in iron-peak elements, with [$\alpha$/Fe] abundance ratios possibly being increased by as much as a full dex \citep{Reddy-et-al-2006-a}. 
Even though they  are enhanced over iron,  
$\alpha$-elements may still have sub-solar abundances, with typical [$\alpha$/H] of -0.3 dex \citep{Zink-et-al-2023-a}.
Under the thick disk hypothesis  3I should thus be old, have a surface modified by long exposure to cosmic radiation, have a less than solar metallicity and smaller than solar abundances  of  iron and iron-peak elements.

\subsection{Planetesimal Formation in the Thick Disk}
\label{subsec:thick disk-star-formation}

While a number of thick disk stars have confirmed planets, \cite{Bashi-Zucker-2022-a} and \cite{Zink-et-al-2023-a} both found reductions in the occurrence rates of rocky planets in thick disk FGK stars. \cite{Hallatt-Lee-2025-a} explain these reductions through UV photoevaporation of protoplanetary disks in dense molecular cloud
``clumps'' thought to exist in the early galaxy 
\citep{Clarke-et-al-2019-a}. These clumps give rise to numerous hot, bright, massive  stars emitting both  Far UV (FUV, between 6 and 13.6 eV, or 206 - 91 nm) and Extreme UV (EUV, between 13.6 and 124 eV, or 91 - 10 nm) photons. Simulations show that 
the FUV flux could have been 4 - 6 orders of magnitude higher in a thick disk star-formation cloud compared to  a solar analogue cloud, and there also could have been a significant EUV flux, which is completely blocked in a solar analogue cloud by the neutral interstellar medium (ISM). 
Unlike solar-type star formation, where most
heating and radiation tends to come from the central object (i.e., the proto-star), in the thick disk case the UV flux could have both irradiated and heated the entire protoplanetary disk, to $\sim$300 K in the ``fiducial'' thick disk cloud of
\cite{Hallatt-Lee-2025-a}. 

If 3I is an object from the thick disk it thus likely formed in an environment very different, and much harsher, than the thin disk star formation regions in the solar neighborhood, and this different birth environment should have left observational signatures on 3I's composition and structure. It would have been in that case likely that 3I was born warm and red, due to heating and irradiation from an intense FUV flux
\citep{Dalle-Ore-et-al-2011-a}. If 3I was created at a mean temperature of order 300 K then  "supervolatile" molecules, such as CO, CO$_{2}$, N$_{2}$ and CH$_{4}$ \citep{Prialnik-et-al-2004-a}, should be largely absent. 
\cite{Seligman-et-al-2022-a} show that, for solar system comets, the C/O ratio and the presence of CO$_{2}$ and CO depend on whether or not the comet forms outside the CO$_{2}$ and CO  ``snow lines,'' the  temperature in the protoplanetary disk where CO$_{2}$ and CO  condense (at 47K and 20K, respectively). This suggests that, if 3I was condensed in a warm thick disk star formation region, it should produce in its coma very little CO and CO$_{2}$.

Whether amorphous ice could have formed under these conditions is unclear, but it seems necessary under the thick disk hypothesis as the  amorphous-crystalline ice phase transition seems to be the only plausible  energy source for  coma creation out at the orbit of Jupiter in the absence of other volatiles \citep{Prialnik-Jewitt-2024-a}. 

\subsection{Polycyclic Aromatic Hydrocarbons}
\label{subsec:PAH-info}

Polycyclic aromatic hydrocarbons (PAH), a class of  organic compounds composed of multiple  aromatic rings, have been found to be common in the ISM, with complicated emission spectra including peaks at 3.3, 6.2, 7.6-7.7, 8.6 and
11.0-11.2 $\mu$m \citep{Hudgins-Allamandola-2004-a}, together with a plethora of weaker lines, spectral plateaus and other  features \citep{Peeters-2011-a}. These spectral features correspond to different vibrational modes in the PAH molecules, and thus provide  information about the types of molecules present \citep{Hrodmarsson-et-al-2025-a}. Specifically, the 3.3 $\mu$m band arises from the C–H stretching mode of neutral PAHs, the 6.2 $\mu$m band is thought to
be generated by aromatic C–C stretching, the 8.6 $\mu$m band
from in-plane C-H bending modes, the 7.6-7.7 $\mu$m band results from a combination of  stretching and  in-plane bending modes, while  the 11.0 and 11.2 $\mu$m emissions correspond to out of plane motions of single hydrogen atoms for  cationic and neutral PAH molecules, respectively
\citep{Stock-et-al-2014-a,Maragkoudakis-et-al-2020-a,Knight-et-al-2022-a}.

PAH band emissions, resulting from molecular vibrations excited by absorption of Far UV photons, are common in our and other galaxies, including in the thick disks of other spiral galaxies \citep{Howk-2012-a}, and
have also been found in emissions from solar system comets \citep{Venkataraman-et-al-2023-a}, the surfaces of Iapetus, Hyperion and Phoebe \citep{Cruikshank-et-al-2014-a,Dalle-Ore-et-al-2011-a}, and the upper atmosphere of Titan 
\citep{Lopez-Puertas-et-al-2013-a}. 
Star formation regions also tend to show strong  PAH IR emissions, with the molecular weight typically increasing  with metallicity, and also with the intensity of the UV radiation in the environment  \citep{Knight-et-al-2022-a} although it is not clear if this is because the radiation facilitates the formation of these molecules, or if the  UV flux preferentially photo-disassociates the lighter aromatic molecules. 

Studies have shown a strong correlation between the metallicities of starburst galaxies and the strength of the PAH emissions in the IR using a transition level of 0.35 Z$_{\bigodot}$ \citep{Engelhardt-et-al-2008-a,Cannon-et-al-2006-a}, where Z$_{\bigodot}$  is the solar  metallicity, while
\cite{Whitcomb-et-al-2024-a} found that the ``total PAH-to-dust luminosity ratio remains relatively constant until reaching a threshold of $\sim$ 2/3 Z$_{\bigodot}$, below which it declines smoothly but rapidly.'' 
\cite{Santos-et-al-2017-a} 
found that the metallicity of the Milky Way thick disk is (0.577 $\pm$ 0.276) Z$_{\bigodot}$. Using the model of \cite{Whitcomb-et-al-2024-a}, the predicted   PAH-to-dust luminosity ratio for a thick disk object would vary between $\sim$2\% and $\sim$15\%.
In addition, in their model the size of the PAH grains increases with radiation hardness (and decreases with metallicity) because PAHs are removed from
the population 
by photo-disassociation
starting from the smallest grains up, and this process becomes more efficient at lower metallicities. 
This raises the possibility that determination of the IR PAH emission and grain size in the 3I coma could potentially provide a tighter constraint on the conditions of its star formation region than is possible through stellar modeling.  

The emission strength ratio of emissions at 6.2 and 7.7 $\mu$m has been used as an indicator of the size of the PAH molecules involved, but the 11.2/3.3 $\mu$m line ratio has been shown to be a much better indicator of PAH molecular size, scaling  with
the number of carbon atoms (N$_{\mathrm{C}}$) \citep{Maragkoudakis-et-al-2020-a}.
Unfortunately, it does not appear that any of the encounter spacecraft can observe both of these spectral bands (see Sections \ref{subsec:martian-close-approach} and \ref{subsec:Juice-perihelion-campaign}) and that only the JWST will be able to simultaneously observe the full PAH spectrum from 3I, although it cannot do this at perihelion due to its solar elongation restrictions (see Section \ref{subsec:observational-limits}). We therefore recommend that JWST acquire spectra of the full (at least 3 - 12 $\mu$m) PAH band in 2025 December, after 3I leaves the JWST solar exclusion zone. 

\section{Observing 3I from Encounter Spacecraft}
\label{sec:3I-Spacecraft-Obs}

As it will not be possible to launch or re-direct a spacecraft to perform a close 3I flyby during its passage through the inner solar system \citep{Yaginuma-et-al-2025-a}, potential observations from space will depend on the location and capabilities of existing instruments on spacecraft already in various heliocentric or planetary orbits. We therefore made a list of relevant operational scientific missions already in space (we know of none that are planned to be launched during the 3I encounter period) and evaluated their different instruments, yielding 16 spacecraft in all: 11 ``encounter'' spacecraft (those targeted at other solar system planets, moons or asteroids), which could conduct \textit{ad hoc} observations although they are each devoted primarily to the exploration of bodies (see Table \ref{table:spacecraft-cameras}), and the 5 solar observatories discussed in Section \ref{subsec:solar-probes}, which can possibly provide monitoring data as part of their regular observations.

Unless one or more spacecraft happen to pass through 3I's tail (see Section \ref{subsec:3I-tail-observing}), 3I  will be observed in the electromagnetic spectrum, and  relevant spacecraft instruments  will be cameras and spectrometers, also listed in Table \ref{table:spacecraft-cameras}. The primary parameters of interest for such observations are the distance, the brightness of the target, the elongation (the Sun-observer-3I angle) and the phase angle (the observer-3I-Sun angle). Table \ref{table:closest-approach} provides these parameters at the time of each encounter spacecraft's closest approach to 3I, using the JPL Horizons dynamical ephemeris service. In order to evaluate potential observations, it is of course also necessary to estimate the brightness of the target, which we did using standard cometary models. 

\subsection{3I Magnitude Estimates}
\label{subsec:Magnitudes}
 
When 3I was first discovered \citep{Bolin-et-al-2025-a}, its absolute magnitude was estimated to be H = 11.84, yielding a fairly large  estimated radius $>$ 10 km, but that estimate clearly included the brightness of the coma at discovery. Within a day of discovery, imaging  revealed cometary activity and  the existence of a coma \citep{Seligman-et-al-2025-a}. Using observations from the Vera C. Rubin Observatory, \cite{Chandler-et-al-2025-a} estimated  H = 13.7 $\pm$ 0.2 mag, giving a  nucleus radius upper bound of (5.6 $\pm$ 0.7) km. 
In this paper, 
we use the absolute visual magnitude lower bound of   
15.4 mag, and the comparable  nucleus radius upper bound of 2.8 km, from \cite{Jewitt-et-al-2025-a} for the 3I nucleus. Unfortunately, none of the encounter spacecraft will be able to resolve a nucleus of that size, even at their closest approaches, and so 3I itself will be a one pixel source for the observations discussed here.

Approximate estimates for comet magnitudes are provided by a simple empirical equation
\citep{Shanklin-2023-a}
\begin{equation}
\mathrm{m}\ =\ \mathrm{H}\ + 5\ 
\log_{10}(\mathrm{D}_{\mathrm{observer}})\ +\ 
\mathrm{K}_{1}\ \log_{10}(\mathrm{R}_{\mathrm{Sun}})
\label{eq:comet_magnitudes}
\end{equation}
where H is the absolute magnitude,  K$_{1}$ 
is the magnitude ``slope parameter,''
D$_{\mathrm{observer}}$ is the distance (in Astronomical Units, or AU) between the observer and the comet,  and R$_{\mathrm{Sun}}$ is the distance (in AU) between the comet and the Sun.

For an unchanging rocky body, such as an asteroid or a comet nucleus, K$_{1}$ = 5, corresponding to a brightness change $\propto$ 1 / R$_{\mathrm{Sun}}^{2}$, is assumed, and so the only free parameter is H, the body's absolute magnitude. Currently, we use H = 15.4 for the 3I nucleus, the lower bound provided from HST observations by \citep{Jewitt-et-al-2025-a}. For the comet's coma we adapted the model estimate for 3I obtained on 2025 Aug 20 from the  Comet Observation database (COBS)  maintained by Crni Vrh Observatory, which provided an estimate of  H$_{\mathrm{coma}}$ = 9.4 and K$_{1}$ = 3.8 based on data from 2025 July 1 through August 18. Results from these models provide the values shown in Figure \ref{fig:Magnitudes-PMJ-C} and Tables \ref{table:closest-approach} through \ref{table:Juice-close-approach}.
Note that, to be conservative, the coma magnitude estimates 
provided in those tables do not include a model for  the brightening likely
with observations at high phase angles; even if the coma does not brighten intrinsically as 3I warms, observations at phase angles $\gtrsim$ 90$\degree$ could easily be 1 or 2 magnitudes brighter than the values provided here. 

\begin{sidewaystable} 
        \centering
        \begin{tabular}{ c  c  c c c } 
          \hline
           Spacecraft  & Camera  & 
           Angular     & Spectral &  References  \\
                       &         &
           Resolution  & Band  &   \\
        
        \hline
Psyche & Imager & 10'' &  7 bands over 439-1015 nm & \citep{Bell-et-al-2025-a} \\
\\
Mars   & MRO HiRISE        & 0.206'' & 3 channels & \citep{McEwen-et-al-2024-a}\\
       &  &  & 400-600, 550-850 and 800-1000 nm\\
        & Tianwen-1 HiRiC  & 0.389'' & panchromatic mode & \citep{Meng-et-al-2021}\\
    & Maven Imaging Ultraviolet Spectrograph (IUVS) & 103''  & 110 - 340 nm & \citep{McClintock-et-al-2015-a}\\
    & Emirates Mars Ultraviolet Spectrometer (EMUS) & 0.18$\degree$ & 100 - 170 nm &\citep{Holsclaw-et-al-2021-a} \\
    & Emirates Mars InfraRed Spectrometer (EMIRS) & 0.32$\degree$ & 6 - 100  $\mu$m & \citep{Edwards-et-al-2021-a} \\
\\
Lucy   & L'LORRI & 1'' &  Panchromatic from 420-795 nm & \citep{Weaver-et-al-2023-a}\\ 
       & L'RALPH MVIC & 5.9'' & 5 bands over 350-950 nm & \citep{Simon-et-al-2025-a}\\
\\
Hera   & Asteroid Framing Cameras & 18.74''
& Panchromatic & \citep{Vincent-et-al-2024-a}\\
\\
Juice  & JANUS  & 2'' & 13 bands over 380-1080 nm & \citep{Palumbo-et-al-2025-a} \\
 & MAJIS & 30'' & Imaging spectrometer over 500-5550 nm  & \citep{Poulet-et-al-2024-a}\\
        &   &  & 4 nm (VIS) and 7 nm (IR) channels  \\
  &  UV imaging Spectrograph (UVS) & 0.04$\degree$& 55-210 nm & \citep{Davis-et-al-2021-a}\\
\\
Osiris Apex & PolyCam   & 2.78'' & Panchromatic 500-800 nm & \citep{Rizk-et-al-2018-a}\\
\\
Juno     & JunoCam      & 130.5'' & 4 bands, 420–520 nm, 500–600 nm,  & \citep{Hansen-et-al-2017-a}\\
 &      & & 600-800 nm, and 880–900 nm & \\
  &  Stellar Reference Unit & 128'' & 1 band, 450-1100 nm & \citep{Becker-et-al-2023-a}\\
        \hline
       \end{tabular}
    \caption{\textbf{Potential Spacecraft Instruments.} 
 Selected instruments possibly available for 3I observations. Note that while the high resolution imagers will generally be able to resolve the 3I coma at closest approach, the various imaging spectrometers generally will not. Unfortunately, none of these instruments would resolve the comet nucleus given the current upper bound of its radius. The Europa Clipper Europa Imaging System (EIS) will not be able to image 3I as its cameras have protective covers that will not be removed until 2027. 
    }
\label{table:spacecraft-cameras}
\end{sidewaystable} 

\begin{table}[ht]
        \centering
        \begin{tabular}{ c  c  c c c c c c c c c} 
          \hline
           Body  & Date of &
           \multicolumn{2}{ c }{3I Distance}  &  \multicolumn{2}{ c }{Magnitude} 
           &  \multicolumn{2}{ c }{Elongation}  & \multicolumn{2}{ c }{Coma Diameter} & Sky\\
                 &  closest approach  &  to Body & to Sun & Nucleus  & Coma & at Body & at Earth & Angular & Pixels & Velocity\\
        \hline
Psyche &  2025 Sep 4 05:00 & 0.301 AU & 2.420 AU & 15.2 & 8.3 & 119.9$\degree$ & 70.26$\degree$ & 114'' & 11 &  0.327 ''s$^{-1}$\\
Mars (HiRISE) &  2025 Oct 3   04:00 & 0.195 AU & 1.663 AU & 13.0 & 6.7 & 126.43$\degree$ &  26.89$\degree$ & 177'' & 859 & 0.611 ''s$^{-1}$\\
(HiRIC) & &  &  &  & & & &  & 455\\
Lucy &  2025 Oct 17 13:00 & 2.140 AU & 1.425 AU & 17.9 & 11.6 &  11.00$\degree$ &  6.10$\degree$ & 16'' & 16 & 0.052 ''s$^{-1}$\\
Hera & 2025 Oct 21  21:00 & 0.984 AU & 1.384 AU & 16.1 & 9.9 &  6.89$\degree$ &  2.77$\degree$ & 35'' & 2 & 0.117 ''s$^{-1}$\\
Europa Clipper & 2025 Oct 28 07:00 & 1.002 AU & 1.359 AU & 16.1 & 9.9 &  3.12$\degree$ &  11.05$\degree$ & 34'' & - & 0.113 ''s$^{-1}$\\
Juice & 2025 Nov 4 10:00 & 0.428 AU & 1.375 AU & 14.3 & 8.1 &  169.36$\degree$ &  22.05$\degree$ & 81'' & 40  & 0.317 ''s$^{-1}$\\ 
Osiris-Apex & 2025 Dec 12 16:00 & 1.521 AU & 2.104 AU & 17.9 & 11.5 &   124.53$\degree$ &   92.45$\degree$  & 23'' & 8  & 0.069 ''s$^{-1}$\\ 
Earth  & 2025 Dec 19 06:00  & 1.796 AU & 2.286 AU & 18.5 & 12.0 &  \multicolumn{2}{ c }{106.95$\degree$} & 19'' & & 0.051 ''s$^{-1}$\\
Juno & 2026 Mar 16 12:00 & 0.354 AU & 5.238 AU & 16.8 & 9.8 &  65.91$\degree$ &  106.37$\degree$ &  97'' & $<$ 1 &  0.257 ''s$^{-1}$\\
          \hline
       \end{tabular}
    \caption{\textbf{Close Approach Opportunities.} 
    Conditions at the closest approach of 3I to various spacecraft (and planets with orbiting spacecraft).  3I is angularly close to the Sun for  Europa Clipper, Hera, and Lucy near the time of their closest approaches; the Europa Clipper will not be able to image 3I and Hera and Lucy would not be able to observe 3I at such low elongations. ``Psyche'' here refers to the spacecraft, not to the minor planet, ``Imager'' to the Psyche Multispectral Imager \citep{Bell-et-al-2025-a}, elongation angles are the Sun-body-3I angle, times are rounded to the nearest hour, and the coma angular size assumes a constant coma diameter of 25,000 km  \citep{de_la_Fuente_Marcos-et-al-2025-a}. Both the coma size and brightness estimates are thus intentionally conservative, assuming no intrinsic enlargement or brightness increases during the encounter period. Sky motions are with respect to background stars and do not include orbital velocities for Juno or the Mars orbiters.
    }
\label{table:closest-approach}
\end{table}

\subsection{Observational Limits Near 3I's Perihelion}
\label{subsec:observational-limits}

The 3I perihelion will be on 2025 October 29 at 11:34:35
$\pm$ 57 s TT, at a solar distance of 1.35639 $\pm$ 0.00010  AU, just inside the perihelion distance of the Martian orbit (1.38 AU); 
Table \ref{table:perihelion-observing} provides observing conditions at time of perihelion. Unfortunately, the perihelion will happen when 3I is almost on the other side of the Sun as seen from Earth, at an elongation of 12.79$\degree$.
Observations from Earth will be especially difficult in the period before the 3I perihelion; the minimum solar elongation for an observer at the terrestrial geocenter is 2.59$\degree$ on 2025 October 21 at $\sim$05:00 UTC, and the terrestrial elongation will be $\le$ 25 $\degree$ for the entire period of October 4 to November 6 of 2025.

\begin{table}[ht]
        \centering
        \begin{tabular}{ c   c c c c r } 
          \hline
           Body  & 
           3I Distance  &  \multicolumn{2}{ c }{Magnitude} 
           & Phase &  \multicolumn{1}{ c }{Elongation}  \\
         &  to Body  & Nucleus & Coma & Angle & \multicolumn{1}{ c }{at Body}   \\
        \hline
Juice & 0.547 AU & 14.8 & 8.6  & 28.35$\degree$  &  135.34$\degree$ \\
Hera &  1.051 AU  & 16.2 & 10.0 & 156.98$\degree$ & 12.90$\degree$\\
Mars &  1.309 AU & 16.7 & 10.5 &  68.77$\degree$ & 57.21$\degree$   \\
Osiris-Apex & 2.173 AU & 17.8 &  11.6  & 10.65$\degree$ & 16.65$\degree$ \\
Lucy &   2.217 AU  & 17.8 & 11.6 & 164.04$\degree$ &  6.02$\degree$   \\
Psyche &  2.277 AU & 17.8 & 11.7 & 63.94$\degree$ & 36.00$\degree$ \\

Earth  & 2.308 AU & 17.9 & 11.7 & 9.34$\degree$ & 12.79$\degree$ \\
Juno & 5.371 AU & 19.8 & 13.6 & 74.98$\degree$ & 14.64$\degree$ \\
          \hline
       \end{tabular}
    \caption{\textbf{Conditions at perihelion.} 
    Observing conditions at  various spacecraft (and planets with orbiting spacecraft) for 3I at its perihelion, 2025 October 29 at $\sim$11:34 TT and 1.3563  AU from the Sun. Juice will clearly be the spacecraft best positioned to observe 3I at perihelion; the 3I solar elongation angle at Earth will be 12.8$\degree$, small enough to make many terrestrial  observations impossible.
    }
\label{table:perihelion-observing}
\end{table}

Clearly, terrestrial observations will be limited at one of the most crucial times for observing 3I.
Terrestrial optical telescopes, which cannot observe faint objects in daytime, will be restricted at best to observing near dawn or dusk. The Rubin / LSST, for example, has a 15$\degree$ elevation angle limit, which (together with the daytime restrictions) prevents it from observing 3I from October 6 to November 14 of 2025, a 39 day period which includes both the 3I perihelion and the Juice close approach. Radio telescopes can observe during daytime but will have elongation limits. The ALMA radio telescope array, for example, has a 15$\degree$ solar elongation limit; it will not be able to observe 3I between October 11 and October 30, a 19 day obscuration period which includes the 3I perihelion. 

Space telescopes typically have stringent elongation limits to avoid heating or damaging the detectors. The James Webb Space Telescope (JWST) has a 85$\degree$ lower elongation limit, which will prevent JWST observations of 3I from 2025 August 25  to 2025 December 09, while the Hubble Space Telescope (HST) has a 55$\degree$ elongation limit, yielding a 3I blackout period from 2025 September 14 to November 23. Note that, with these limits, neither telescope will be able to observe 3I at the time of either the Mars or Juice close approaches, or the 3I perihelion. 

X-ray spectroscopy can and should be attempted as X-ray observations can provide useful insight to 3I's volatile content, searching for components such as H$_{2}$, N$_{2}$ and CO \citep{Cabot-et-al-2023-a,Maltagliati-2023-a}.
X-ray telescopes also have elongation limits comparable to the large space optical telescopes, with the elongation limit for the Chandra X-ray telescope being 46.4$\degree$, and for XMM-Newton 70$\degree$ \citep{Cabot-et-al-2023-a}, these limits corresponding to 3I blackouts of 2025 September 20 - November 19, and 2025 September 4 - December 1, respectively. 
These constraints will limit X-ray observations of 3I volatile production during the perihelion passage period.

Most spacecraft in deep space also have limits to observing at low solar elongations, although specific limits can be hard to find in the literature. (The solar probes and solar monitors in heliocentric orbits can observe very close to the Sun with coronagraphs and other specialized instruments; their observations are considered in Section \ref{subsec:solar-probes}.)
Calibration work with the  Lucy L'Lorri telescope has shown that its sensitivity drops rapidly as the solar elongation of the target goes below 90$\degree$ due to scattered light entering the optics \citep{Stern-at-al-2025-a}.
Undoubtedly, there will also be hard limits in how close spacecraft operators will allow their instruments to be pointed at the Sun; here, 
based on known spacecraft (and terrestrial) limits, we assume here that these will be roughly 45$\degree$. 

\begin{figure*}[ht!]
\plottwo{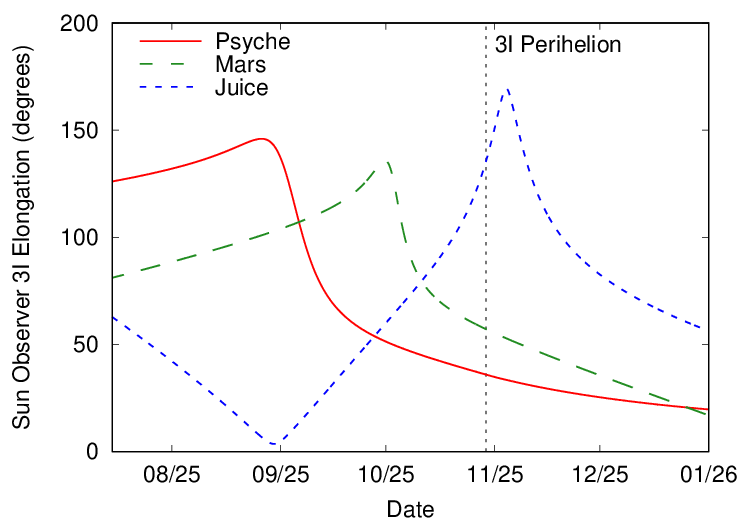}{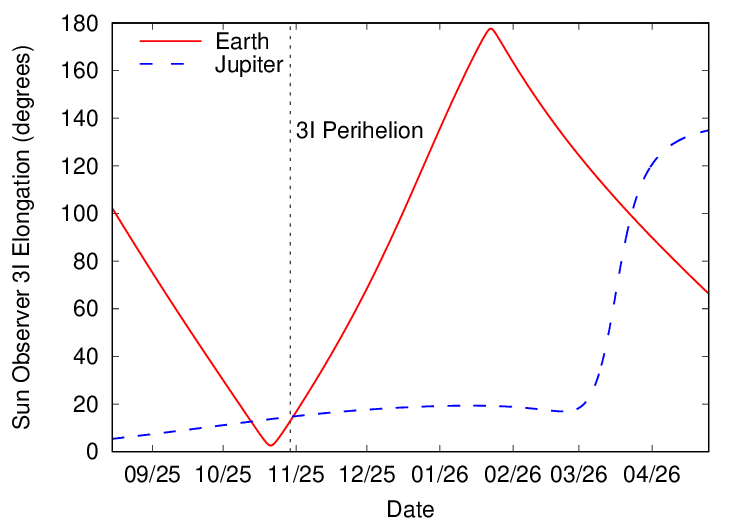}
\caption{Elongations (the 3I-Observer-Sun angle) for the Psyche spacecraft, observers on or near Mars, and the Juice spacecraft (left image), and for observers on or near the Earth and Jupiter (right image). Note that both spacecraft and terrestrial telescopes will have trouble observing 3I at small elongations, as described further in the text in Section 
\ref{subsec:observational-limits}. 
\label{fig:Elongations-1-2}}
\end{figure*}

\section{Potential 3I Spacecraft Observations} \label{sec:spacecraft-exploration}

For any potential spacecraft observations of an \textit{ad hoc} target such as 3I it is necessary both to determine when observations are possible, and, if they are possible, whether they are compelling, or if they could be performed better from Earth or some other vantage point. Comets have notoriously unpredictable behavior near perihelion, and spacecraft monitoring of 3I when it is hard to observe or unobservable  from Earth will be especially valuable, but observations from different vantage points, and at different wavelengths, can also provide otherwise unavailable data.

For any comet nearing perihelion, repeated observation of its light curve (the change in brightness with time), its phase curve (the change in brightness as a function of the Observer-Comet-Sun angle) and of any outbursts or fragmentation events are of scientific interest, as is regular astrometry (measurements of its position), and compositional spectroscopy. It should be possible, with the available instruments,  to determine whether any rocky material is silicate or carbonaceous, to determine its water content, and the amount of any low-temperature volatiles. The available UV spectrometers should be able to sense Lyman alpha emissions from the 3I coma, which is important in constraining the total hydrogen content. Of course, as we have never encountered a thick disk  object before, it is likely that it will surprise us, and so observations from a variety of instruments and viewpoints could yield unanticipated results.

\subsection{3I Astrometry}
\label{subsec:astrometry}

The dynamical ephemeris of any interstellar object is limited by the  astrometric observations performed in its passage through the solar system. Although the rate of 3I astrometric observations has been relatively large, with a mean of over 60 observations acquired per day since its discovery, its rapid passage through the solar system, and the constraints on terrestrial observations during its perihelion passage, will limit the final ephemeris accuracy obtained for 3I's orbit, especially if observations are only available from near-Earth space. 

As an active comet, 3I should experience some level of non-gravitational acceleration due to outgassing \citep{Hui-Jewitt-2017-a}; determining this acceleration, which can be used to constrain both the mass and the volatile emission rate of the ISO, is an important goal for 3I science. This will be  especially difficult if the 3I nucleus is as large as its current upper bounds. With a radius of 2.8 km, based on other cometary data, its non-gravitational acceleration should be order 3 $\times$ 10$^{-13}$ AU  day$^{-2}$
\citep{Seligman-et-al-2023-a}. Such a small acceleration  would not have been detected with the astrometric data collected for either 1I or 2I; these  have non-gravitational  
formal errors of  $\sim$10$^{-8}$ AU day$^{-2}$
and  $\sim$10$^{-9}$ AU day$^{-2}$, respectively.

The accuracy needed to detect non-gravitational accelerations of 3I (3 $\times$ 10$^{-13}$ AU  day$^{-2}$ $\sim$ 6 km yr$^{-2}$), probably cannot be obtained solely from terrestrial (and near-Earth) angular astrometry. In the absence of planetary radar (which will not be possible for 3I given the orbital geometry of this encounter) the distance to the body must be inferred from angular measurements, which, if it is only done from one location, introduces a geometrical dilution of precision in the range, and thus degrades the accuracy of the estimated orbital parameters. Fortunately, imagery from interplanetary spacecraft can be used to improve our knowledge of the 3I orbit. 

Spacecraft astrometry of 3I can be
done with long-exposure optical navigation (OpNav) images which will also be useful for coma dust imaging.
\textit{In situ} OpNav astrometry of 3I at the arc second accuracy level (routinely done in many deep space missions) at spacecraft-3I distances $\lesssim$ 1 AU should, when combined with near simultaneous observations from Earth, enable the direct determination of the 3-D position of  3I to within a few hundred km in a single  observation. Repeated such observations (which do not have to be all performed by the same spacecraft) should substantially improve the 3I orbit determination. Stellar occultations, if possible (and they are being attempted by a variety of groups) should provide determinations of the angular position of 3I with an accuracy of 10 milli arcseconds (mas) or better
\citep{Ferreira-et-al-2022-a}, equivalent to 10 km (or better) positioning in the plane of the sky. Occultations would also provide a direct determination of the size and shape of the 3I nucleus, information that will not be available by any other means. A combination of terrestrial and encounter spacecraft astrometry, together with repeated occultation measurements, should be able to detect a non-gravitational acceleration, if it exists, at the 10$^{-13}$ AU day$^{-2}$ level.

\subsection{Coma Dust Phase Angle Measurements}
\label{subsec:Coma-Dust-Phase-Angles}

Comparison of cometary coma images from different locations allows for the phase angle response of the coma, a means of determining the  size, size distribution, shape, and composition of dust
particles plus information on their columnar density along the line of sight. The Rosetta mission to comet 67P  
\citep{Bertini-et-al-2017-a} and theoretical modeling  show a ``half-U'' shaped phase angle function, with a low plateau at low phase angles (i.e., the backwards scattering case as the Sun-3I-observer angle approaches 0$\degree$), a relatively sharp transition at phase angles of order 90$\degree$, and a high plateau, with up to two orders of magnitude increase in brightness in the extreme forward scattering case, as the  Sun-3I-observer  angle goes to 180$\degree$ \citep{Keiser-et-al-2024-a,Bertini-et-al-2025-a}.

The Psyche close approach, in particular, will occur at times when observations will be possible from Earth and from other spacecraft, enabling near-simultaneous sampling of the phase angle function at a variety of different observing geometries. This is discussed further in Section \ref{subsec:psyche-phase-measurements}.

\subsection{Comet Coma Studies}
\label{subsec:coma-study}

Information about the composition of small celestial bodies largely comes from either gas spectra, or the colors of solid bodies (including dust in comet comas). At present, there are no reports of detections of emission lines from the 3I coma, but the infrared colors of the coma dust have already been used to deduce the present of water ice in the coma \citep{Yang-et-al-2025-a} (a similar search did not work in the case of 2I \citep{Yang-et-al-2019-a}.
While emission lines were never detected from 1I, and it is still unclear whether that ISO was actually active, a variety of spectral lines were detected from 2I, including C$_{2}$, NH$_{2}$ and CN, with 2I being nitrogen rich \citep{Deam-et-al-2025-a}
and carbon-chain depleted \citep{Kareta-et-al-2019-a}. Obviously, this work should be repeated with 3I.

Gaseous atomic nickel was found in the coma of 2I with the X-shooter spectrograph of the Very Large telescope 
\citep{Guzik-et-al-2021-a} at a distance of 2.322 AU from the Sun, a distance that
3I will pass on 2025 September 07 (inbound) and 2025 December 20,  (outbound), in both cases with reasonable terrestrial viewing restrictions. Since both Ni I and Fe I lines
have been detected in the comas of solar system comets \citep{Manfroid-et-al-2021-a}, and since the abundance of both nickel and iron should be less  in thick disk $\alpha$-element ISOs, attempts should be made to determine the existence of both Ni I and Fe I line emissions in the 3I coma.

The light curve of a comet provides information about the volatiles creating the comet's coma. Water ice will start to sublimate at $\sim$ 3 to 2.6 AU from the Sun (i.e., starting 2025 August 16 to 29 and ending 2025 December 20 to 2026 January 11 for 3I). Figure \ref{fig:Magnitudes-PMJ-C} shows that, at least in principle, encounter spacecraft (especially Juice and the Mars spacecraft) could continue the coma light curve throughout the period without observations for the Earth.

\subsection{Observation of Comet Fragmentation}

An important goal of spacecraft observations near perihelion will be to observe and characterize any cometary fragmentation of 3I. 
The interstellar object 2I split and released a small sub-component roughly between 2020 Mar 12th and Mar 20th \citep{Bolin-et-al-2020-a,Jewitt-et-al-2020-a}, roughly 3 months after its perihelion when it was $\sim$3 AU from the Sun, a distance 3I  reached on 2025 August 15 as it approaches perihelion and will reach again on  2026 January 11 as it leaves the solar system. 

Fragmentation of a comet such as 3I is thus possible (but of course not inevitable) at any time during its transit of the inner solar system. This is of course an excellent argument for regular 
observations of 3I by spacecraft, both  by encounter spacecrafts and, if possible, by solar probes, during  the periods the ISO can  best or only be observed via spacecraft.

\subsection{Observational Implications of a Thick Disk ISO Origin}
\label{subsec:Observational_Implications}

Thin and thick disk stars and ISOs can be defined kinematically (as in Section \ref{sec:dynamical-origins} here) or chemically \citep{Vieira-et-al-2022-a}; it seems clear that these definitions do not entirely overlap.
The upcoming observations of 3I offer the opportunity to test its origin chemically, and thus confirm or deny the kinematic conclusions about its origin. If 3I is from the thick disk, observations should provide insight into the star and planet-forming processes operating early in the history of our galaxy. 

The compositional influences of a formation in , the thick disk includes a low metallicity, decreased fraction of iron, and an increased water content
\citep{Santos-et-al-2017-a,Weeks-et-al-2025-a,Lintott-et-al-2022-a}. It also likely includes formation at a fairly high temperature in an irradiated environment. An early history in a warm, wet, environment should mean that super-volatiles such as N$_{2}$ or CO are largely lacking from the 3I coma. It also could mean that there are minerals on 3I, such as hydrous  phyllosilicates, more typically associated with asteroidal or even planetary surfaces. The existing interplanetary spacecraft armada is largely intended to study asteroids, rocky planets or icy moons, and is well equipped to determine the surface properties of grains in the coma, and also the type of outgassing 3I is experiencing.

Thick disk ISOs should be relatively sparse compared to ISOs from the thin disk and it may be a relatively long time before another one transits the inner solar system. 
\cite{Eubanks-et-al-2021-a}
used the velocity distribution of stars from the Gaia Early Data Release
3 Catalogue of Nearby Stars to estimate that 
$\sim$6\% of incoming ISOs should come  from the thick disk, while
\cite{Anguiano-et-al-2020-a} estimated that the local (kinematically selected) thick-to-thin-disk stellar number density  is 2.1\% $\pm$ 0.2\%.  

These estimates imply that order 10 - 30  additional ISOs will have be found to ensure a 50\% chance of finding another thick disk object, assuming that stellar systems in both disks produce on average the same numbers of ISOs per star. Even if the Rubin LSST increases the 
ISO discovery rate to $\sim$1 year$^{-1}$ it could still take one or more  decades before another such object is found.

\section{Observing Campaigns}
\label{sec:observing-campaigns}

\begin{figure*}[ht!]
\plotone{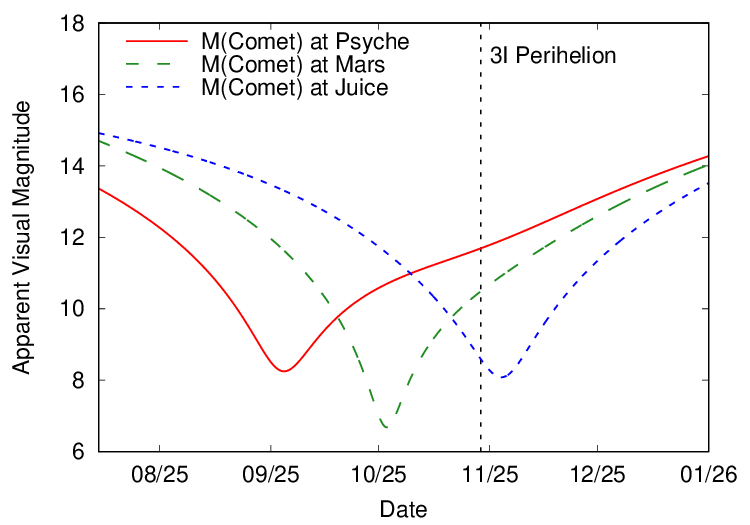}
\caption{3I cometary coma magnitude estimates from Equation \ref{eq:comet_magnitudes}  for observations of 3I from the Psyche and Juice spacecraft, and also for the spacecraft on or near Mars. 
\label{fig:Magnitudes-PMJ-C}}
\end{figure*}

With the available orbital ephemerides, magnitude estimates and observational limits the three relatively close approaches where we recommend spacecraft observations are around the Psyche, Mars and Juice encounters, centered on 2025 September 4, October 3, and  November 4, respectively. Figure \ref{fig:Magnitudes-PMJ-C} shows that, if the resources are available, the coma is predicted to be  brighter than 10th magnitude for at least one of these  three spacecraft in the period September through November in 2025; repeated spacecraft observations during this period would be valuable if possible.  

The encounter spacecraft discussed have in general a high resolution camera and also an imaging spectrometer, or spectrometers. (There also may be other instruments neglected here.) The high resolution cameras have spectral filters (``bands'' in Table  \ref{table:spacecraft-cameras}) intended to distinguish between different geological units, or ices versus solid rocks, which will be useful in distinguishing different solid components in cometary coma dust. The imaging spectrometers have higher spectral resolution, and will be able to distinguish different emission lines in the coma spectrum. 
While the larger terrestrial telescopes of course can have larger apertures and higher resolutions than these spacecraft units, Psyche will be 8.53 times closer to 3I than the Earth at its closest approach, Mars 12.78 times closer, and of course, Earth observations will be impossible or severely limited during the 3I perihelion and the Juice close approach 6 days later.

3I has a very high velocity in its passage through the solar system, leading to the large sky velocities shown in Table \ref{table:closest-approach}. These angular velocities exceed 0.3 ''s$^{-1}$ for Psyche and Juice, and 0.6 ''s$^{-1}$ for spacecraft at Mars, indicating that the spacecraft will have to either roll to slew with the motion of 3I, or adopt some form of high dynamic range synthetic tracking  \citep{Zhai-et-al-2018-a} where many short exposures are shifted, stacked, and combined to achieve longer exposure images without streaking.

The elongation limits discussed in  \ref{subsec:Magnitudes} indicate that the Lucy and Hera spacecraft, which have 3I passing between them and the Sun during its perihelion passage, will probably not be able to image 3I at their closest approaches. 
However, if they can image 3I in early September, observations at the larger phase angles their orbits will provide then could be important in characterizing the 3I coma dust.
Unfortunately the Europa Clipper EIS will not be able to image 3I at all in 2025; we removed this spacecraft from the subsequent observation tables.
In addition, as is described in Section  \ref{subsec:3I-tail-observing}, the Europa Clipper, Hera, and even the Lucy spacecraft may possibly be able to sample the 3I cometary tail.

OSIRIS-APEX will be  relatively close to Earth at the time of the 3I perihelion, and thus will suffer from the same observational constraints that terrestrial and other near terrestrial telescopes will have. However, in mid December, when both it and the Earth have their closest approaches, observations will be possible and will offer a different vantage point (and a useful astrometric parallax) from the observations at Earth.

The Juno spacecraft orbiting Jupiter will not be able to usefully observe 3I until 2026 March (see Figure \ref{fig:Elongations-1-2}). The observational possibilities for Juno are different enough that they are out of scope for this paper and will have to be discussed elsewhere.

\subsection{A Psyche Close Approach Phase Measurement Campaign}
\label{subsec:psyche-phase-measurements}

The spacecraft Psyche will have a relatively close approach to 3I on 2025 September 4, at a distance of 0.302 AU. Psyche is currently on a trajectory between its launch in 2023 October, and a planned Mars gravity assist in 2026 May. As Psyche would nominally  be under thrust from its solar-electric propulsion system during the encounter period, and this may have to be temporarily stopped to obtain imagery,  it is not likely that a long sequence of observations can be obtained. 

The Psyche encounter will occur when most terrestrial telescopes, the HST, and Lucy, Hera,  and the Mars orbiting spacecraft can all observe 3I.
 (Note that the full Moon on 2025 September 7th may  limit some terrestrial visual and IR observing during the Psyche close approach period.)
 Observations from Earth and from the Psyche spacecraft alone could provide a good determination of the 3I phase function at phase angles of 53.9$\degree$ and 23.1$\degree$, with observations from Mars providing an intermediate point at 38.8$\degree$
and observations from Lucy, if they can be performed at an solar elongation of 46.6$\degree$, extending it to 72.1$\degree$.
The Juice spacecraft will have just finished a Venus gravity assist (2025 August 31), will have  a very small solar elongation and won't be able to observe 3I near this date, while the Hera  spacecraft could provide phase angles similar to the Psyche observations. Of course, even if Psyche or other spacecrafts cannot observe on these specific dates, observations on other nearby dates would undoubtedly extend the observational distribution of the 3I coma phase angle   During its entire period of observability, 3I phase angles from Earth only extend to a maximum of 30.58$\degree$. Clearly, coordinated observations around  the time of the Psyche close approach offer a good opportunity to the phase angle distribution curve and better characterize the nature of the 3I coma dust during its passage through the solar system.

\begin{figure*}[ht!]
\plotone{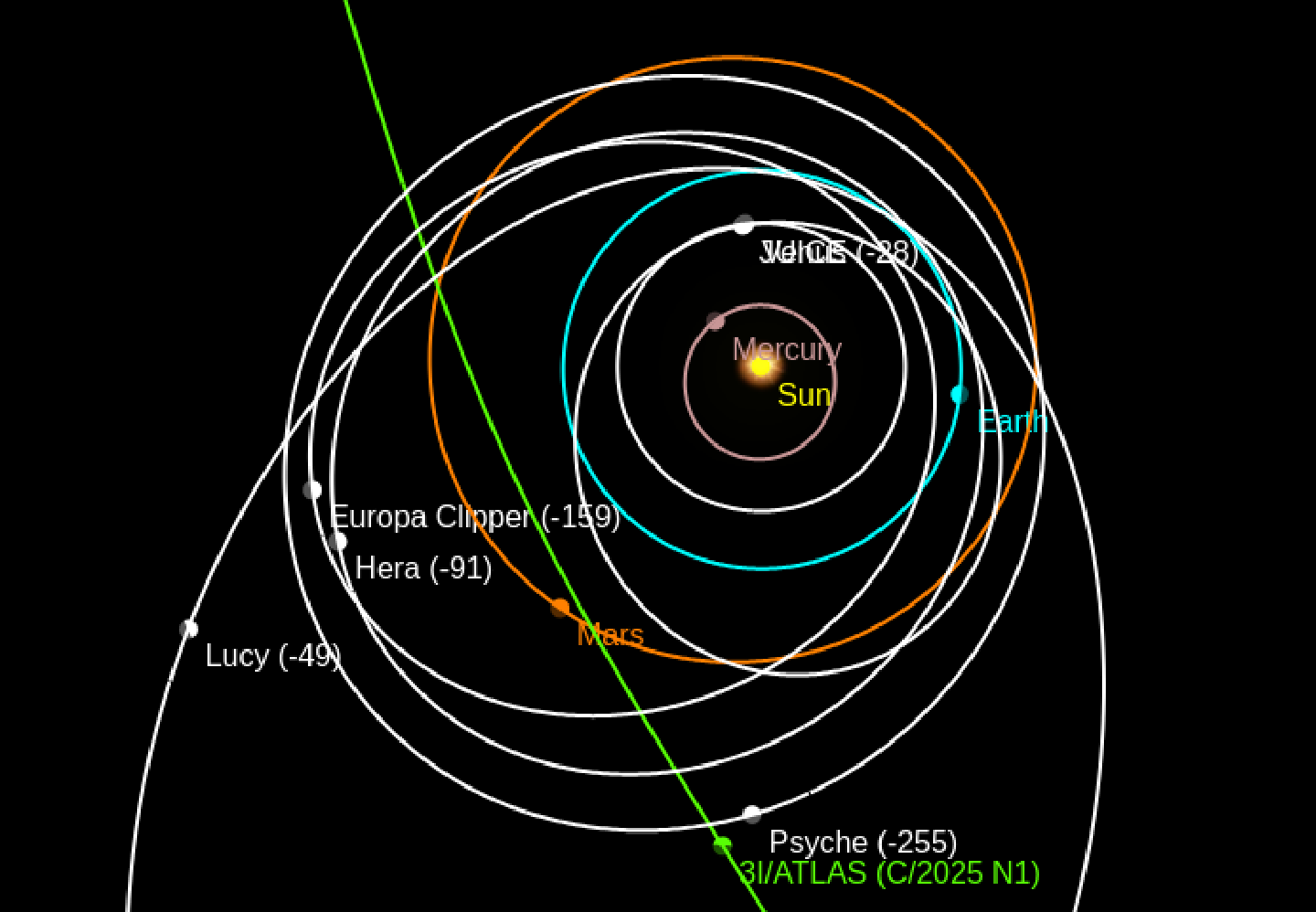}
\caption{A North polar view of the inner solar system on 2025 September 4, at the closest approach of the Psyche spacecraft to 3I. Note that in this view from the North, the retrograde 3I is moving to the upper left, while the planets are orbiting counter-clockwise and the Psyche spacecraft is therefore opposing 3I's motion, with a total relative velocity of 71.58 km s$^{-1}$ 
At this time, the Psyche Imager, together with telescopes on or near the Earth and Mars, and two of the three exterior spacecraft (Lucy and Hera) could potentially observe 3I, providing data at up to 5 different phase angles. 
(The solution for this figure and Figures \ref{fig:Mars_close_approach} and \ref{fig:Juice_close_approach} were provided by Tony Dunn of \url{http://orbitsimulator.com/}.)
\label{fig:Psyche_close_approach}}
\end{figure*}

\begin{table}[ht]
        \centering
        \begin{tabular}{ c   c c r r } 
          \hline
           Observer  & 
           3I Distance  &  Comet 
           &  Elongation & Phase  \\
         &  to Observer  & Magnitude  & Angle & Angle  \\
        \hline
Psyche & 0.301 AU & 8.3 & 119.88$\degree$ &  53.92$\degree$ \\
Mars & 1.456 AU & 11.7 & 105.88$\degree$ &  38.77$\degree$ \\
Hera & 2.461 AU & 12.8 & 61.04$\degree$ &  56.13$\degree$ \\
Earth & 2.567 AU & 12.9 & 70.24$\degree$ &  23.09$\degree$ \\
Osiris-Apex & 2.636 AU & 13.0 & 66.63$\degree$ &  22.89$\degree$ \\
Lucy & 2.927 AU & 13.2 & 46.55$\degree$ &  72.07$\degree$ \\
Juice & 3.119 AU & 13.3 & 9.91$\degree$ &  2.91$\degree$ \\
       \end{tabular}
    \caption{\textbf{Observing at the closest approach to the Psyche spacecraft.} 
    Observing conditions at  various spacecraft (and planets with orbiting spacecraft) for 3I at the time of its closest approach to the spacecraft Psyche, 2025 September 4 at 05:00 (see also Figure \ref{fig:Psyche_close_approach}). 
    }
\label{table:Psyche-observing}
\end{table}

\subsection{The Martian Close Approach Campaign}
\label{subsec:martian-close-approach}

\begin{figure*}[ht!]
\plotone{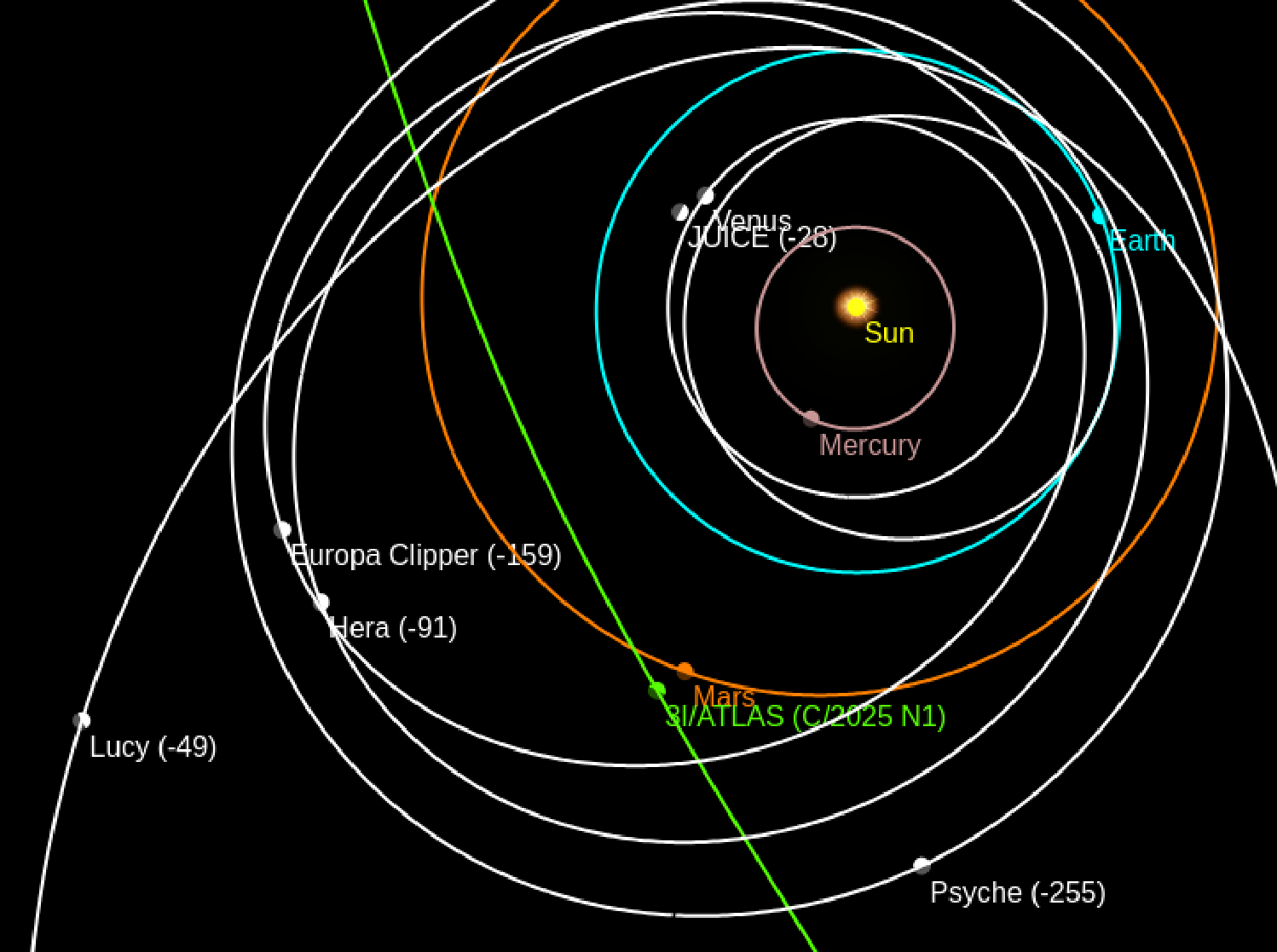}
\caption{A North polar view on 2025 October 3 at the closest approach of Mars (and its orbiting spacecraft) to 3I. Mars and the ISO have opposing velocities, with a total relative velocity of 86.15 km s$^{-1}$, and a motion, relative to the fixed stars, in the Martian sky at closest approach of 0.61'' s$^{-1}$. 
\label{fig:Mars_close_approach}}
\end{figure*}

The closest approach to 3I (Figure \ref{fig:Mars_close_approach} and Table \ref{table:closest-approach}) will be by the various spacecraft at or near Mars, on 2025 October 3rd, at a distance of 0.195 AU, or 29 million km. The U.S. Mars Reconnaissance Orbiter (MRO) and the Chinese Tianwen-1 orbiter, with their high resolution imagers, and the Emirates Hope orbiter Emirates Mars Ultraviolet Spectrometer (EMUS) 
\citep{Holsclaw-et-al-2021-a}, will provide detailed imagery at a relatively close range. 

The High Resolution Imaging Science Experiment (HiRISE) camera \citep{McEwen-et-al-2024-a} is a 0.5 m reflecting telescope with a resolution of 1 microradian (206 mas), yielding a resolution of $\sim$30 km at closest approach to 3I, equivalent to resolution of a 6.4 m telescope at Earth. 
The Tianwen-1 HiRIC camera is an off-axis three-mirror astigmatic (TMA) optical system with an aperture of 0.387 m \citep{Meng-et-al-2021} and a resolution of 1.887 microradian (389 mas), equivalent to a 5 meter telescope at Earth. As  neither the HST and JWST will be able to observe 3I at this time, these telescopes in Mars orbit should yield better resolution than  visual telescopic resolution from the Earth's surface. Since the vantage points of Earth and Mars are separated by roughly 45$\degree$, the combination of Mars orbiting and Earth based images will enable the 3-dimensional (3-D) modeling of the entire coma, and highly accurate 3-D positioning of the coma, and the nucleus, if it can be distinguished in the coma at the closest approach.

\begin{table}[ht]
        \centering
        \begin{tabular}{ c   c c r r } 
          \hline
           Observer  & 
           3I Distance  &  Comet 
           &  Elongation & Phase  \\
         &  to Observer  & Magnitude  & Angle & Angle  \\
        \hline
Mars & 0.195 AU & 6.7 & 126.43$\degree$ &  48.17$\degree$ \\
Psyche & 1.231 AU & 10.7 & 49.84$\degree$ &  95.70$\degree$ \\
Hera & 1.336 AU & 10.9 & 44.09$\degree$ &  101.92$\degree$ \\
Juice & 1.847 AU & 11.6 & 64.14$\degree$ &  24.39$\degree$ \\
Lucy & 2.245 AU & 12.0 & 25.71$\degree$ & 118.44$\degree$ \\
Osiris-Apex & 2.465 AU & 12.2 & 26.99$\degree$ &  15.30$\degree$ \\
Earth & 2.493 AU & 12.2 & 26.88$\degree$ &  15.78$\degree$ \\
       \end{tabular}
    \caption{\textbf{Observing at the closest approach to Mars.} 
    Observing conditions at  various spacecraft (and planets with orbiting spacecraft) for 3I at the time of its closest approach to the planet Mars, 2025 October 3 04:00 (see also Figure \ref{fig:Mars_close_approach}). Psyche, Hera, and Juice could all provide useful supporting observations for this encounter, especially in determination of the coma's phase angle brightness variation. 
    }
\label{table:Mars-observing}
\end{table}

\subsubsection{FUV Observations at 3I's Close Approach}
\label{subsubsec:EUV}

If 3I is a water-rich body, it will sublimate and outgas more water (and other hydrogen containing material) as it moves closer to the Sun. This hydrogen in the coma will be excited by solar UV radiation, creating Lyman-$\alpha$ and Lyman-$\beta$
spectral lines, at  121.6 nm and 102.6 nm, respectively. The two UV spectrometers currently orbiting Mars, the Emirates Mars Ultraviolet Spectrometer (EMUS) on the Hope spacecraft \citep{Holsclaw-et-al-2021-a} and the 
Imaging Ultraviolet Spectrograph (IUVS) on the Maven spacecraft \citep{McClintock-et-al-2015-a} could both potentially observe the 3I Lyman-$\alpha$ emission, while the EMUS has the wavelength range to also observe the Lyman-$\beta$ line. 

In October of 2014, the Oort cloud comet 
C/2013 A1 (Siding Spring) made a very close approach to the planet Mars. The Maven IUVS detected the  Lyman-$\alpha$ coma of the comet on 2014 October 14, with an integration time of 60 s per pixel and at a distance of 0.1709 AU, close to the closest approach distance of 3I to Mars (0.195 AU)
\citep{Crismani-et-al-2015-a}. From these observations, they deduced a  water production rate
of (1.1 $\pm$ 0.5) $\times$ 10$^{28}$ molecules s$^{-1}$.

As the same telescope could be used to observe the EUV coma of 3I at almost the same distance, it presumably would be able to detect its coma if its water production  rate is comparable to the C/2013 A1 water production rate. The first detection of water activity from 3I came from observations on 2025 July 31 - August 1 with the 
Neil Gehrels-Swift Observatory and deduced a water production rate of (1.35 $\pm$ 0.27) $\times$ 10$^{27}$ molecules s$^{-1}$
at a distance of 3.81 AU \citep{Xing-et-al-2025-a}, or 
$\sim$ 12\% of the rate at 229\% of the distance to the Sun. While it is difficult to predict the future water production rate, Q(r), of any comet, models of the form 
\begin{equation}
\mathrm{Q}(\mathrm{r})\ =\ \mathrm{A}\ \mathrm{r}^{\mathrm{B}}
\label{eq:water-activity}
\end{equation}
where A is a constant, r is the heliocentric distance in AU, and B the activity gradient (production rate slope), which can vary from -4 at 3 AU to -2.5 at 2 AU \citep{Marshall-et-al-2019-a}.

If the mean B for 3I is $\lesssim$ -2.6  between 3.8 AU and 1.66 AU, it should be possible to observe the 3I Lyman-$\alpha$ coma from using the IUVS. This would provide a direct measurement of the comet's hydrogen production rate, and possibly those of other elements, if those lines can be detected. 

\subsubsection{IR Emissions from Mars Orbit}
\label{subsubsec:IR-at-Mars}

The   Emirates Mars InfraRed
Spectrometer (EMIRS) instrument \citep{Edwards-et-al-2021-a}, a 17.78-cm diameter Cassegrain telescope, may offer the best opportunity for detecting  water IR emissions from 3I. This spectrometer can sense the strong 6.1 $\mu$m water spectral line, which 
results from the water molecule H–O–H bending mode and thus specifically indicates the presence of water (due to its strength this line cannot be observed astronomically from even the driest places on Earth), and which is outside of the bandwidth of the JUICE/MAJIS spectrometer. (The JWST Mid-Infrared Instrument (MIRI) can also observe this line, but will not be able to do so between 2025 August 25 and 2025 December 9, when 3I will  be 2.7 and 2.0 AU from the Sun, respectively.) 

The EMIRS observes wavelengths from 6 - 100 $\mu$m, which means it cannot observe the PAH band at 3.3 $\mu$m, and thus cannot determine the size of any PAH molecules observed through the 3.3/11.2 $\mu$m band ratio (see the discussion in Section \ref{subsec:PAH-info}). The available band ratio of 6.2/7.7 $\mu$m is not reliable in this regard, but the 6.2/11.2 $\mu$m band ratio may be available and is sensitive to the charge of the PAH molecules. If PAH IR emissions can be detected by EMIRS, it would also be good to arrange coordinated observations with the JUICE MAJIS, which might enable 3.3/11.2 $\mu$m band ratio observations.  

\subsubsection{Observations from the Surface of Mars}
\label{subsubsec:surface-of-Mars-obs}

\begin{figure*}[ht!]
\plotone{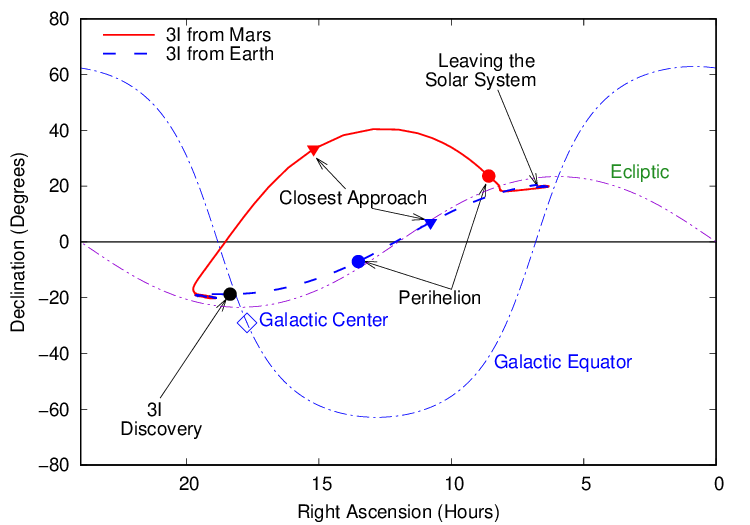}
\caption{A plot of 3I trajectory across the sky for both the Earth and Mars. 3I never gets that close to Earth, and its orbit is near the ecliptic, and so its sky track stays there too. 3I does get close to Mars, its orbit is more inclined relative to the orbit of Mars and so its track  goes high in the northern martian sky. Since the Perseverance mars rover is located at 18$\degree$ North latitude, it will have a better view of 3I at the time of closest approach. 
\label{fig:Mars-sky-tracks}}
\end{figure*}

Figure \ref{fig:Mars-sky-tracks} shows the track of 3I in the terrestrial and martian skies.
At latitudes of -4$\degree$ and +18$degree$, respectively, the Curiosity and Perseverance rovers could both observe 3I at night  from the surface of Mars during the period when it was closest to the planet. The diminishing 3I solar elongation would not prevent surface observations until January of 2026; more stringent limits will be set by the brightness of 3I and the available surface camera systems. 

The two currently operational Mars rovers both carry Mastcam-100 cameras with a 100 mm focal length and a resolution of 74 microradians, or 15.25'', per pixel. On October 3rd at closest approach 3I will be at a declination of 33.355$\degree$, and the rotation of the martian sky at that declination will be 12.212'' s$^{-1}$ (the sky motion of 3I at closest approach will be 0.611 '' s$^{-1}$), implying that the longest exposure shouldn't be longer than $\sim$1 s to avoid motion smearing. The limiting magnitude with that exposure would be $\sim$ 8, which would limit the observability of 3I (in the standard brightness model) to the period September 28 to October 11 of 2025. Apparently vibrations prevent the Mastcams from mechanically tracking moving objects, so  synthetic tracking 
\citep{Zhai-et-al-2018-a} would have to be implemented to extend the limiting magnitude to 10 or higher and track 3I into perihelion. 

The purpose of these observations would be general monitoring of changes in the brightness of the coma, the size of the coma (which should be well resolved by the Mastcam-100 throughout this period) and a search for fragmentation events near perihelion. 

\subsection{The Juice Perihelion Campaign}
\label{subsec:Juice-perihelion-campaign}

Juice will come within 0.428 AU of 3I on 2025 November 4 at a relative velocity of 98.41970 km$^{-1}$, and 5.24 times closer than the Earth at that instant. The 3I perihelion is 6 days earlier, on October 29, and the minimum 3I elongation at Earth, 2.59$\degree$, is 8 days before that, on October 21.  
Tables \ref{table:perihelion-observing} and \ref{table:Juice-close-approach} indicate that only Juice and the spacecraft at Mars will be able to closely observe 3I during its perihelion, and Juice will have by any measure the best view then. 

Although 3I will pass close to the Sun from the perspective of Earth at its superior conjunction, Juice will not; the Juice elongation at Earth will never be less than 18.15$\degree$ during the entire passage of 3I through the solar system.  Communications with the Earth should not be interrupted by the solar plasma, potentially enabling nearly immediate transmission of 3I data back to Earth. 

JANUS, an anastigmatic catadioptric telescope, is the major Juice camera system, with an aperture of 115 mm and a field of view of 1.72$\degree$ $\times$ 1.29$\degree$ \citep{Palumbo-et-al-2025-a}, or 1.92 $\times$ 1.44 million km projected to 3I's distance at closest approach, which may  not be sufficient to
span an extended comet tail in a single image. At closest approach to 3I, the JANUS resolution is 620 km while the HiRISE resolution (at 1.456 AU) is $\sim$2100 km, 3.4 times worse, and HiRISE observations at solar elongations of 52.81$\degree$ may not be possible. 

\subsubsection{3I Visual and IR Spectrometry at Perihelion}
\label{subsubsec:PAH-at-Perihelion}

The Juice Moons And Jupiter Imaging Spectrometer (MAJIS) is an imaging spectrometer operating in the
visible and near-infrared spectral range which is intended to ``investigate upper atmospheric
chemistry and exogenic inputs from the stratosphere to the thermosphere'' on Jupiter, and the H$_{2}$O exosphere of Ganymede \citep{Plainaki-et-al-2020-a}, and could perform the same investigation on the coma of 3I. MAJIS covers wavelengths
 from 0.50 to 5.55 $\mu$m in two spectral bands, 0.5 to 2.3 $\mu$m and 2.3 to 5.55 $\mu$m, with
a spectral resolution of 4 nm  and 7 nm, respectively. 

MAJIS could allow for a number of interesting investigations of the gas content of the 3I coma, including neutral gases such as CO$_{2}$ at 4.23 $\mu$m and 2.7 $\mu$m, H$_{2}$O at 2.7 $\mu$m, O$_{2}$  at
1.27 $\mu$m, and CO at 4.6 $\mu$m, and a wide range excited molecules. With an angular resolution of 30'', the MAJIS will have a transverse resolution of $\sim$9000 km at closest approach, potentially enabling mapping of gas emissions across the  3I coma.

The longer wavelength MAJIS band covers the 3.3 $\mu$m PAH spectral line region, which results from CH stretch vibrations.
Detection and monitoring of PAH IR emissions during the 3I perihelion passage is an important science goal for this encounter period, and the MAJIS appears to be the best instrument to observe this line emission during the 3I perihelion passage. Coordinated observations with the Hope spacecraft EMIRS (Section \ref{subsubsec:IR-at-Mars}) could potentially allow for determinations of the full suite of PAH IR emissions, from 3.3 to 11.2 $\mu$m, possibly around October 21, when the distances to 3I are roughly equal (at 0.9 AU) and the solar elongations are a reasonable 64$\degree$ (for Mars) and 103$\degree$ (for Juice).

\begin{figure*}[ht!]
\plotone{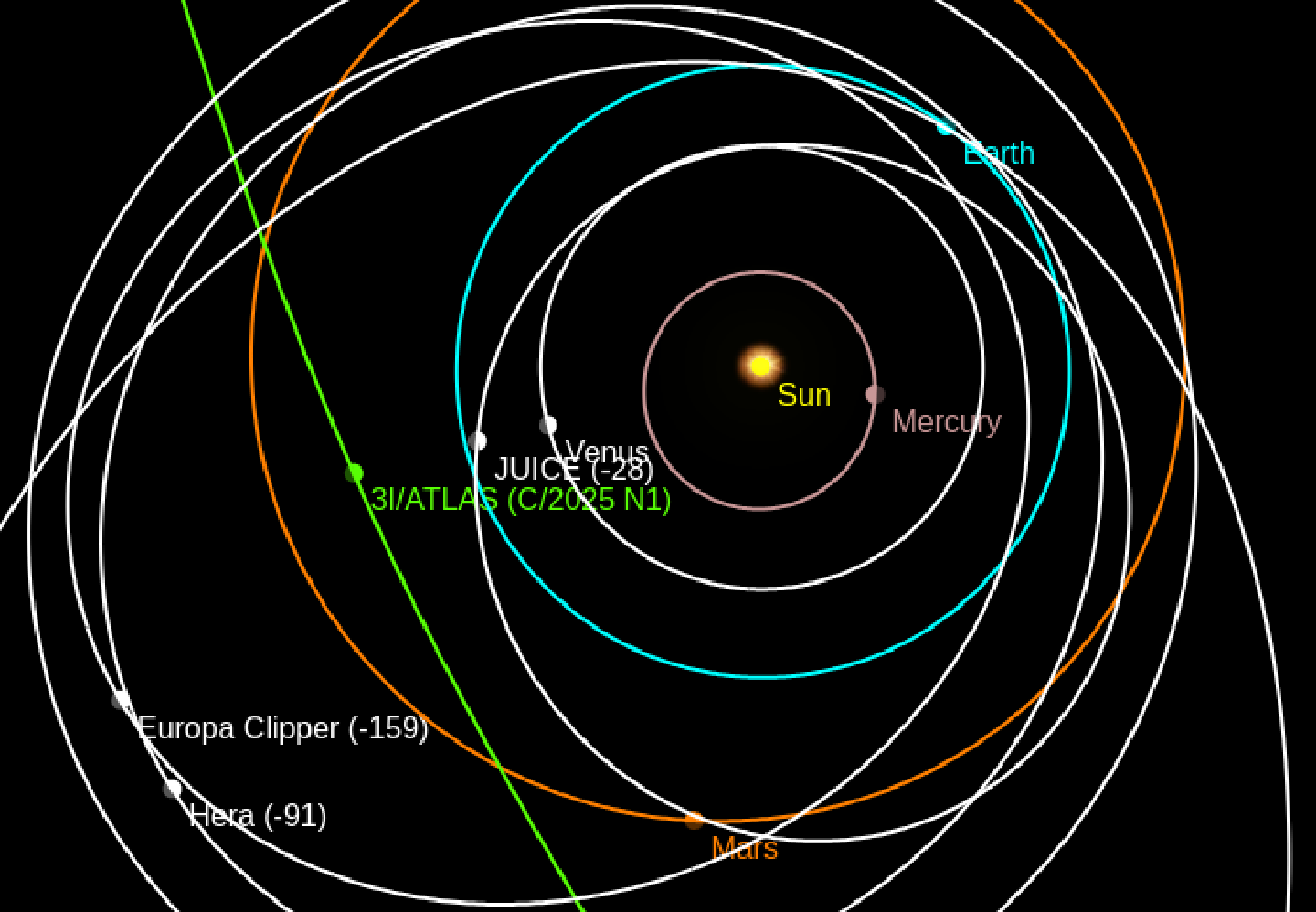}
\caption{A North polar view on 2025 November 4 at the closest approach of the Juice spacecraft to 3I, 6 days after the 3I perihelion. With the Juice elongation angle at Earth being $\gtrsim$18.15$\degree$ throughout the 3I perihelion and Juice encounter period, the Earth should be able to communicate with the spacecraft during this period, although optical observations of 3I will likely be impeded. 
\label{fig:Juice_close_approach}}
\end{figure*}

\begin{table}[ht]
        \centering
        \begin{tabular}{ c   c c r r } 
          \hline
           Observer  & 
           3I Distance  &  Comet 
           &  Elongation & Phase  \\
         &  to Observer  & Magnitude  & Angle & Angle  \\
        \hline
Juice & 0.428 AU & 8.1 & 169.36$\degree$ &  7.35$\degree$ \\
Hera & 1.186 AU & 10.3 & 24.52$\degree$ & 134.51$\degree$ \\
Mars & 1.589 AU & 10.9 & 52.81$\degree$ &  60.19 $\degree$ \\
Osiris-APEX & 2.076 AU & 11.5 & 27.84$\degree$ & 16.98$\degree$ \\
Earth & 2.243 AU & 11.7 & 22.06$\degree$ &  15.71$\degree$ \\
Lucy & 2.311 AU & 11.7 & 10.66$\degree$ &  151.23$\degree$ \\
Psyche & 2.506 AU & 11.9 & 33.45$\degree$ & 51.82$\degree$ \\
       \end{tabular}
    \caption{\textbf{Observing at the closest approach to Juice.} 
    Observing conditions at  various spacecraft (and planets with orbiting spacecraft) for 3I at the time of its closest approach to the spacecraft Juice, 2025 November 4 at 10:00 (see also Figure \ref{fig:Juice_close_approach}). Except for the spacecraft at Mars, all the other encounter spacecraft have small solar elongations and are unlikely to be able to observe 3I near this date.  
    }
\label{table:Juice-close-approach}
\end{table}

\subsubsection{EUV and FUV Spectrometry at Perihelion}
\label{subsubsec:UVS-at-Perihelion}

The JUICE ultraviolet spectrograph (UVS)
is the fifth in a series of imaging UV spectrometers that include instruments that have already flown,
Rosetta-Alice, New Horizons-Alice, LRO-LAMP, and Juno-UVS \citep{Davis-et-al-2024-a,Davis-et-al-2021-a}. This UVS covers a wavelength range of 50-204 nm with a spectral resolution of 0.5 - 1.5 nm, depending on wavelength and position in the sky, with a FOV of 7.3$\degree$ $\times$ 0.1$\degree$ and a best-case spatial resolution of 0.07$\degree$, or 75,000 km at closest approach. The Juice UVS is thus unlikely to resolve the 3I coma, but could image a 3I cometary tail, if this tail is bright enough for that signal to be above the noise. 

The Juice UVS should be able to  observe most of the same spectral lines as the EMUS, with 
Hydrogen Lyman-$\alpha$ (121.6 nm) and Oxygen O I (130 nm) likely providing the strongest signals. With its wider bandwidth, the UVS could potentially observe some harder spectral lines that EMUS cannot, such as O II at 83.4 nm and Ar I at 86.7 nm, if these are sufficiently excited at perihelion. 

\subsection{In Situ Observations of the 3I Comet Tail}
\label{subsec:3I-tail-observing}

As it does not seem possible to redirect the trajectory of any spacecraft to flyby 3I,  the best possibility of directly sampling its material would be if one or more spacecraft passes through the 3I tail. Comets can develop 3 types of tails, plasma (or ionized gas), dust and neutral tails, with the plasma tail shining by stellar excitation of ionized gas, and the others, by reflection of sunlight. Even for 3I, with a heliocentric velocity at perihelion of 68.33 km s$^{-1}$, the solar wind is much faster (typically 400 km s$^{-1}$ or faster), and supersonic, so that ion tails are typically radially aligned in the anti-Sun direction. The dust in dust tails is partially dominated by radiation pressure, and so streams away from Sun but trails behind the plasma tail. Finally, larger particles ejected from the comet's surface at relatively low velocities are not strongly influenced by solar radiation pressure and so stay on or near the comets orbit, strung out behind it and forming both the neutral tail (also sometimes call the anti-tail) and also potentially forming meteor showers if the Earth, or another planet, passes through the comet's orbit. 

As Figures \ref{fig:Mars_close_approach} and \ref{fig:Juice_close_approach} indicate, the spacecraft Hera, Lucy and Europa Clipper will pass near or through the 3I plasma tail, roughly on October 24, 26 and 28 of 2025, respectively, and maybe also  the dust tail a few days or weeks  later, i at a distance of $\sim$1 AU from the comet's nucleus and $\sim$8 million km out of the comet's orbital plane for Hera and Europa Clipper, and twice that for Lucy. In the 1996 encounter of the ion tail of the comet C/1996 B2 (Hyakutake) 
the material encountered had traveled 3.38 AU away from the comet with a travel time of a week
\citep{Jones-2002-a}, so a tail passage at a distance of 1 or 2 AU is not out of the question, but depends on the exact shape of the tail at the encounter time. 

While the Hera and Lucy spacecraft do not appear to carry any instruments suitable for the direct analysis of comet tail material, the Europa Clipper has two instruments, the Europa Clipper Magnetometer (ECM) \citep{Kivelson-et-al-2023-a} and  Plasma Instrument for Magnetic Sounding (PIMS) 
\citep{Westlake-et-al-2023-a}, which could be used to analyze the ionized material in the plasma tail.

Note that even if the 3I tails do not
appear to reach out to a length of 1 AU in imaging
these spacecraft could still be immersed in material from  the comet, as happened in the case of Comet Hyakutake.
The presence of an observable cometary tail means that there is enough material to detect the emissions of ionized gas, or the reflection of light off the dust lost by the comet; sensitive instruments might be able to detect this material \textit{in situ} even if it cannot be seen from a distance. 

Although long period comets can have meteor showers  from heavier particles distributed along the comet's trajectory (the ``neutral tail'') \citep{Jenniskens-et-al-2021-a}, 
3I will never repeat its trajectory, and so encounters with its neutral tail material would require passage through its trajectory relatively soon after 3I does. Unfortunately, this appears unlikely, as the orbital phasing of the best situated spacecraft (Europa Clipper, Hera, and the ones at Mars) is not favorable.  
Mars will pass relatively close to the 3I trajectory on 2025 September 16, 17 days \textbf{before} 3I passes through that location, and will not come close again to the 3I orbit until  2027 March 3.
The Europa Clipper will come close to the 3I trajectory on 2026 May 6, 7 months and 11 days after 3I passes that particular spot, and Hera will do the same 4 days later. All of these intervals are
sufficiently long that detection of  lingering  material seems unlikely. 

\subsection{Searching for Possible 3I Satellites}
\label{subsec:3I-Satellites}

3I has been in a much more quiet dynamical environment than even a ``new'' comet from the Oort cloud, which only reach the inner solar system after cumulative perturbations by galactic tides and  interactions from stellar close approaches 
\citep{Rickman-et-al-2012-a}. 3I of course is also subject to these influences, but it has presumably spent much of its time in the lower stellar density of the thick disk, where the chances of close stellar interactions are lower. Using the mean free path of \cite{Ye-et-al-2017-a}, 3I has probably never come closer than 
200 AU to any star in a lifetime of 10$^{10}$ yr. 

If 3I has a radius of 2.8 km and a density of 1000 kg m$^{-3}$, it has 
a mass $\sim$ 9 $\times$ 10$^{13}$ kg. 
The Hill spheres of even such a relatively  small object can be quite large in a galactic orbit,  $\sim$0.9 AU from galactic tides, and order 5 $\times$ 10$^{-4}$ AU  (7 $\times$ 10$^{4}$ km) induced by a 1 solar mass star at a distance of 200 AU, a rough estimate of the closest likely stellar approach after several billion years passing through the thin disk. Potentially, 3I could be surrounded by a debris cloud, material ejected from its surface by meteorite impacts, with an extent ranging from almost a lunar distance to almost  an AU. 

The size of small satellites of a body such as 3I in its galactic orbit will be limited by drag from the 
ISM 
\citep{Eubanks-2019-b,Ye-et-al-2017-a}. Here, 
we typical thin disk
ISM properties, in particular a density of 2 $\times$
10$^{5}$ hydrogen atoms m$^{-3}$ (the density of the Local Interstellar Cloud, or LIC \citep{Linksy-et-al-2022-a}), and spherical satellites with a density of 1000 kg m$^{-3}$. Although the ISM drag is very low, so would be the orbital velocities of  small 3I satellites, of order 1 cm s$^{-1}$ to no more than $\sim$3 m s$^{-1}$. With this ISM and satellite density, surviving satellites would need to be at least 10s of meters in diameter to have survived in orbit until the present, and should be easily detected as 3I passes through the inner solar system, even as solar tides (and the expected non-gravitational accelerations)  strip any such satellites away from the 3I system. 

If 3I possesses some sort of orbiting debris cloud with an effective Hill sphere radius of order 1 AU,  material impacting the martian atmosphere might be visible fron the surface as meteors or even meteor showers coming from the 3I radiant at  19.6779746 hours R.A. and -19.050743$\degree$ declination. 
These interstellar meteors could potentially be searched for with the
Curiosity and Perseverance rovers in the period between September 13 and October 23 of 2025 (when Mars is $<$ 1 AU from 3I). 

As orbital velocities for material orbiting 3I at a distance of 1 AU are order 1 mm s$^{-1}$, material lost from the outer reaches of its debris cloud would slowly drift away from 3I, forming a larger unbound debris cloud. By analogy with the wide false stellar binaries found in the Milky Way
\citep{Kamdar-et-al-2019-a}, this unbound debris cloud could be several AU across. Some of that material might thus reach the Earth (or impact the Moon) as interstellar meteors from the 3I radiant, probably peaking in the period 2025 November 23  to 2026 January 8, when Earth-Moon system is within 2 AU of 3I. 

\subsection{Monitoring by Solar Probes}
\label{subsec:solar-probes}

There are several spacecraft, including the solar monitors SOHO, STEREO-A, the four NASA PUNCH satelltes, and the NOAA GOES-19 satellite, together with the Parker Solar Probe (PSP) and the Solar Orbiter, which image regions close to the Sun and thus which could potentially observe 3I. As the trajectory of 3I is close to the ecliptic and as it will pass close to the Sun as seen from the Earth, it could pass through the field of view of these spacecraft. 
At present it does not appear that it will be possible for the Solar Orbiter or the STEREO-A probe to observe the ISO 3I within their FOVs. The real question for the remaining spacecraft is, will 3I be active enough, and thus bright enough, to be visible with the instruments on these spacecraft.

The Parker Solar Probe (PSP)  will come within 1.612 AU of 3I on 2025 October 2. In principle, the PSP will be able to observe 3I with the Wide-Field Imager for Solar PRobe (WISPR) \citep{Vourlidas-et-al-2016-a}  
from late September through early November in 2025, which includes the 3I perihelion, with the solar elongation for 3I at the PSP dropping from 141$\degree$ on September 20th to 16$\degree$ on November 10th. The 3I coma visual magnitude during this period is predicted to be brightest, at 11.2,  on 2025 October 13. Although  WISPR did detect the dust trail of (3200) Phaethon at a visual magnitude of $\sim$ 16.1 $\pm$ 0.3 pixel$^{-1}$
\citep{Battams-et-al-2022-a},  observations of a point-like source are more difficult than observations of extended sources  \citep{Battams-et-al-2020-a}, and will be even more difficult at low elongations, as that will shorten exposure times. The WISPR detection of 3I, although possible, will depend on the comet's activity and extent during the period when it is observable. 
These
 observations will of course also be subject to the operational needs of the spacecraft, and thus will at best only happen sporadically, with the observing plan for these exceptionally challenging observations, under difficult observing and operational conditions, still being worked out. 

The ISO 3I will be within the field of view (FOV) of the C3 camera of the 
Large Angle and Spectrometric COronagraph (LASCO) instrument of the SOHO spacecraft from 2025 October 16 - 26, just before the 3I perihelion passage. The closest angular approach of the comet to the Sun from SOHO will be within $\sim$9.5 solar radii on 2025 October 21; 3I  will thus not be close enough to the Sun to  enter the FOV of the C2 instrument. (Note that these numbers are approximate as JPL Horizon does not presently predict the spacecraft's motion on these dates.) 
Unfortunately, the predicted coma brightness at SOHO during this period is magnitude 11.8, larger than the $\sim$ C3 magnitude limit of 8. While 3I will thus transit across the LASCO C3 field of view, it is unlikely that it will be able to detect the 3I coma, unless its brightness  exceeds expectations. 

 3I will cross the field of view of the CCOR-1 coronagraph on the NOAA GOES-19 satellite on October 9 through November 1. However, the mission-critical nature of CCOR-1 as a purely space weather operations-driven instrument preclude the possibility of a dedicated observing campaign for the comet. The observations by default have sensitivity down to approximately a visual magnitude of 10, so this again would  require a substantial increase in brightness of 3I for any positive detections to be made.

 \subsection{Multi-Year Observations of the 3I Nucleus}
 \label{subsec:multi-year-observations}

As 3I recedes from the solar system its cometary activity should stop, enabling the direct determination of both the albedo and effective diameter of its nucleus, through a comparison of its IR and visual flux. This also will also enable an accurate determination of the nuclear rotation period, and thus possibly its moments of inertia,  by determining its change from  outgassing during its passage close to the Sun. If the 3I nucleus is in the brighter end of its absolute magnitude range it may also be possible, by measuring the decline of the coma magnitude with time as 3I cools, to determine the nature of volatile components still present on and near its surface after perihelion passage.

After 2025 the 3I oppositions (as seen from Earth) will occur near the end of each subsequent calendar year. On December 31st of 2026 and 2027
 3I  will be at a distance of 13.93 and 26.27  AU, respectively, yielding nuclear magnitude estimates of 27.0 - 34.7 and 29.7 - 37.4  for those oppositions. If the 3I nucleus is near the upper end of its size range, both the 2026 and 2027 oppositions  should be observable by the HST and JWST space telescopes, and also possibly extremely large terrestrial telescopes.  

 \section{Conclusions} 
 \label{sec:conclusions}

Observations from the Earth, and the HST, JWST and other space telescopes, will certainly be crucial for determining  the nature of 3I, but its passage through the solar system nearly in the ecliptic plane, but on the opposite side of the Sun as seen from the Earth, offers both the opportunity and the need to develop a coordinated observing campaign from interplanetary spacecraft in deep space.

It appears that 
3I potentially formed in the galactic thick disk; its study thus offers the opportunity to investigate star formation and planetary formation conditions in the early history of the galaxy, from 12 to 9 billion years ago. It may be decades before another thick disk candidate ISO is found; every reasonable effort should be made to take advantage of the scientific opportunities currently being offered our latest interstellar visitor. 

A thick disk 3I should have a relative lack of iron-peak elements and a relative abundance of $\alpha$-elements. Attempts should be made to confirm or deny this spectroscopically. It would probably have come from a warm and highly irradiated protoplanetary disk; this could be confirmed by observations of a relative lack of supervolatiles and from IR spectral signatures of PAH outgassing. 
In particular, we predict that this comet will produce very little CO, and will have a C/O ratio $\lesssim$ 0.6.
The combination of spacecraft and terrestrial observations of the 3I coma should thus enable direct testing of the thick disk hypothesis.

Terrestrial observations from Earth will be difficult or impossible roughly from early October through the first week of November, 2025.
Figure \ref{fig:Elongations-1-2}, and Tables \ref{table:spacecraft-cameras} and \ref{table:closest-approach} show that the observational burden during this period will, to the extent that they can observe, largely fall on the Psyche and Juice spacecraft and the 
armada of spacecraft on and orbiting Mars. 
Our recommendation is that attempts should be made to acquire  imagery from encounter spacecraft during the entire period of the passage of 3I through the inner solar system, and in particular from the period in October and November of 2025, when  observations from Earth and the space telescopes will be limited by 3I's passage behind the Sun from those vantage points.  

A combination of terrestrial and spacecraft imagery could directly determine the 3I coma dust phase angle function, and thus improve our knowledge of the size distribution of thick disk interstellar dust. This would benefit from observations by multiple spacecraft around the time of the close encounter by the Psyche spacecraft in early 2025 September.

Spacecraft observations have, in combination with  terrestrial and near-Earth astrometry,  the potential for greatly improving  determinations of the 3I orbit in the solar system, and thus for  detecting or constraining non-gravitational accelerations, 
by providing a very long baseline for parallactic determinations  of the range to 3I, and also by observing its motions when terrestrial telescopes cannot. If these observations can be made together with  observations of stellar occultations of 3I from the Earth it should be possible to determine non-gravitational accelerations of 3I to a few km yr$^{-2}$.

Spacecraft observations from the period when 3I is obscured by the Sun from Earth can monitor the change in brightness and size of the 3I coma near perihelion, possible fragmentation events near perihelion, and IR observations of at least some its water lines, and also of the PAH spectral lines. FUV and EUV observations may be able to see the Hydrogen Lyman-$\alpha$ and other UV spectral lines in the 3I coma, providing direct constraints on its composition. If the coma is sufficiently bright, the PSP and SOHO solar monitoring spacecraft could also provide useful day-by-day observations of the comet when it is is their fields of view. It is possible that 3I has satellites, or is surrounded by a debris cloud, observations from space may be able to constrain these possibilities. As 3I recedes from the Sun, the space telescopes near the Earth can continue to monitor it, and directly observe its nucleus, as it coma disappears  in the years after its passage through the inner solar system.

The Gaia spacecraft has produced a wealth of data on existing and previously unknown stellar streams, including streams in the thick disk and the galactic halo
\citep{Bonaca-et-al-2025-a}. It will be very interesting if 3I can be associated with a thick disk stream or even a halo stream, which would provide a direct tie between a specific part of the galactic history and the data obtained from 3I. 

Fate and chance have offered us a very interesting third interstellar object, however, on an orbit where some crucial observations will be difficult from Earth.
It may be a long time before another such object, a fairly large comet possibly from the galactic thick disk, is available for study. 
Fortunately, there are sufficient spacecraft on missions of interplanetary exploration to make up for the lack of terrestrial observations at the crucial time of 3I's perihelion passage. In this, and in future encounters with interstellar objects in the solar system, we recommend that the opportunities offered by interplanetary spacecraft for interstellar object exploration be seized to the extent possible. 


\begin{acknowledgments}
We would like to thank Tony Dunn of 
 \url{orbitsimulator.com} for assistance with orbital simulations, and spacecraft scientists for dealing with our numerous questions.
 We would also like to thank Karl Battams, Jim Bell, Lindy Elkins-Tanton, Mark Linton, Alfred McEwen, Carol Polanskey and Olivier Witasse
 for useful discussions, information, and the pointing out of errors. 
\end{acknowledgments}

\begin{contribution}

All authors contributed equally to this interstellar object collaboration. TME prepared and submitted the text.


\end{contribution}

%

\software{find\_orb:
          \url{https://www.projectpluto.com/find_orb.htm}
          }



\bibliography{3I}{}
\bibliographystyle{aasjournalv7}



\end{document}